\documentclass[final,leqno]{siamltex}

\usepackage{amsmath,amsfonts,cite}

\def\bx{\tilde x}

\def\fc{\mathbb{F}}

\def\dx{{\rm d}x}
\def\dy{{\rm d}y}
\def\dbx{{\rm d}\bar x}
\def\tPs{\Check\Psi}
\def\tPh{\Check\Phi}
\def\tph{\check\phi}
\def\ps1{\psi_{\perp}}
\def\en{n^{\rm eq}}
\def\Phk{\Phi^{(k)}}
\def\phjk{\phi_j^{(k)}}
\def\F{\mathbb{F}}

\newtheorem{remark}[theorem]{\textit{Remark}}

\title{Bose-Einstein condensation at finite temperatures:\\
Mean field laws with periodic microstructure}

\author{Dionisios Margetis\thanks{Department of Mathematics, and Institute for Physical Science and Technology,
        and Center for Scientific Computation and Mathematical Modeling,
         University of Maryland, College Park, MD 20742-4015 ({\tt dio@math.umd.edu}). This author's research was partially supported by NSF DMS1517162 at the University of Maryland; and by a Research and Scholarship Award (RASA) of the University of Maryland in the spring of 2014.}}

\begin{document}

\maketitle

\begin{abstract}
At finite temperatures below the phase transition point, the Bose-Einstein condensation, the macroscopic occupation of a single quantum state by particles of integer spin, is not complete. In the
language of superfluid helium, this means that the superfluid coexists with the normal fluid.
Our goal is to describe this coexistence in trapped, dilute atomic gases with repulsive interactions via mean field laws that account for a {\em spatially varying} particle interaction strength. By starting with the $N$-body Hamiltonian, $N\gg 1$, we formally derive a system of coupled, nonlinear evolution equations in $3+1$ dimensions  for the following quantities:
(i) the wave function of the macroscopically occupied state; and (ii) the single-particle wave functions of thermally excited states. 
For stationary (bound) states and a scattering length with {\em periodic microstructure} of subscale $\epsilon$, we heuristically extract effective equations of
motion via periodic homogenization up to second order in $\epsilon$. \looseness=-1
\end{abstract}



\pagestyle{myheadings}
\thispagestyle{plain}
\markboth{D. MARGETIS}{BOSE-EINSTEIN CONDENSATION AT FINITE TEMPERATURE}

%
\section{Introduction}
\label{sec:intro}
%

The {\em macroscopic occupation} of  a single quantum state, known as {\em Bose-Einstein condensation (BEC)}, was first observed in trapped, dilute atomic gases at very low temperatures 
almost two decades ago~\cite{Cornell95,Ketterle95}.  This advance has paved the way
to the precise control of the quantum behavior of  ultracold atomic systems; for broad reviews on BEC, see, e.g.,~\cite{Cornelletal02,Ketterle02,Carusottoetal13,Baoetal13,Carretero08,Pitaevskii03}.

From the theoretical viewpoint, three remarkable features of dilute atomic gases in current experimental settings are the following.
(i) The {\em weak} atomic interactions. In many experiments, these interactions are short ranged, characterized by an effective parameter,
the scattering length $a$~\cite{HuangYang57}; for repulsive interactions, 
$a>0$. (ii) The {\em trapping} external potential. This is needed in order to keep the atoms together. Because of this trap, there is no translation invariance of the Boson system.
(iii) {\em Finite} but ``small'' temperatures below the phase transition point. The actual temperatures of the experiments are very low,
in the range of nano\-degrees Kelvin~\cite{Cornelletal02,Ketterle02}. Nonetheless, the BEC is not complete: particles at the macroscopic quantum state, the {\em condensate}, coexist with many particles that occupy thermally excited states.
In the language
of superfluid helium, the superfluid and the normal fluid
are both present~\cite{London}. 

The first feature [(i)], weak particle interactions, enables a systematic theoretical treatment of the Boson gas, as this is carried out in the insightful works by Lee, Huang and Yang~\cite{leeetal,leeyang58} who modeled BEC in the periodic case; in this setting,
the condensate is the zero-momentum state and thermally excited states correspond to nonzero momenta. In connection to the second feature [(ii)], external potential, an ingenious many-body quantum-mechanical treatment of BEC for systems without translation symmetry
was developed by Wu~\cite{Wu61}. These treatments~\cite{leeetal,leeyang58,Wu61} make use of physically transparent approximations for the many-body Hamiltonian.

In this article, we focus on the the third feature [(iii)], finite temperatures, with a  spatially varying scattering length. Our analysis addresses the following question. What mean field laws are consistent with the many-body quantum dynamics of the trapped Boson gas at finite temperatures, sufficiently below the phase transition? In particular, we carry out the following tasks. 

\begin{itemize}

\item We formally derive, directly from a microscopic model by perturbation of nonrelativistic quantum fields, mean field {\em evolution laws} for the trapped dilute atomic gas at finite temperatures when the scattering length is {\em spatially varying}. Our analysis combines heuristic techniques adopted from~\cite{leeetal,leeyang58,Wu61,Wu98}. The emerging description consists of a system of {\em coupled}, nonlinear Schr\"odinger-type partial differential equations (PDEs) in 3+1 dimensions for: (i)  the wave function,
$\tPh(t,x)$, of the condensate; and (ii) the one-particle wave functions, $\tph_j(t,x)$, of the
thermally excited states. This system of PDEs has {\em spatially-varying coefficients} of 
nonlinear terms.

\item For {\em stationary (bound) states}, we provide an effective description for this PDE system when the atomic interaction has a
{\em periodic microstructure} of subscale $\epsilon$. This $\epsilon$ expresses the ratio of the length over which the scattering length, $a$, varies to
the particle correlation length. We apply classical homogenization theory to 
semilinear elliptic equations in the spirit of Bensoussan, Lions, and Papanicolaou~\cite{hom-book}, inspired by work of Fibich, Sivan and Weinstein on the focusing nonlinear Schr\"odinger equation~\cite{Fibich06}.

\end{itemize}

In regard to the derivation of the mean field laws, the proof of convergence of the high-dimensional microscopic dynamics to low-dimensional PDEs lies beyond our present scope. We invoke an uncontrolled yet physically motivated ansatz for the many-body Schr\"odinger state vector.
Furthermore, the convergence of the two-scale expansion in the periodic homogenization program is not pursued here. Approximate solutions of the PDE system are intended to be the subject of future work. \looseness=-1

\subsection{Motivation}
\label{subsec:motiv}
Our work is motivated by experimental advances in controlling properties of ultracold, trapped atomic systems. Evidently, Bose-Einstein condensates form systems of considerable promise
for precision metrology which may offer
improvement over conventional interferometry~\cite{Ketterle02,Croninetal09,Hagley99}. Atomic gases  undergoing BEC can also be used to probe or emulate properties of condensed-matter systems; e.g., the presence of quantum phase transitions such as the transition between
a superfluid and a Mott insulator~\cite{Morschetal06}. In these applications, the short-ranged atomic interactions are known to play an prominent role.
These interactions can be tuned externally, e.g., by optical means near a
Feshbach resonance~\cite{chin10,Cornish00,Stenger99}. \looseness=-1

These ongoing experimental advances suggest the need to study effects intimately connected to the atomic interactions in non-translation-invariant settings.  A key theoretical aspect of this problem is the possible effect on the description of the Bose-Einstein condensate of a spatially varying scattering length, $a$. This possibility is the subject
of the formal analysis  in~\cite{Margetis12}, in which mean-field and beyond-mean-field implications for the condensate are studied at zero temperature.
The present article is meant to be a first step toward extending the perturbative analysis of~\cite{Margetis12} to the case with finite temperatures, when thermally excited states influence the condensate. \looseness=-1

\subsection{Overview: Mean field limit at zero temperature}
\label{subsec:rev-mf}
It is useful to recall that a system of $N$ interacting Bosons evolving at zero temperature is described by a symmetric wave function obeying the $N$-body Schr\"odinger equation. This 
description requires the use of $3N$ spatial coordinates and, thus, is deemed as unwieldy for physical prediction if $N\gg 1$. It is desirable to reduce the many-body evolution to PDEs for variables defined in lower dimensions. A related variable is the condensate wave function, which typically lives in 3 spatial dimensions. 

For repulsive atomic interactions, this reduction usually leads to a mean field law: a defocusing, cubic nonlinear Schr\"odinger equation, herein referred to as the Gross-Pitaevskii-Wu equation (GPWE)~\cite{Gross61,Pitaevskii61,Wu61}. In this PDE, the coefficient of the nonlinear term is proportional to the scattering length, $a$; cf. section~\ref{subsec:mf-intro}. It has been rigorously shown by Elgart, Erd\H{o}s, Schlein and Yau~\cite{Elgart06,Erdos06,Erdos07} how the reduced, one-particle evolution, including the scattering length, properly emerges as $N\to\infty$ from the many-body dynamics by use of the Bogoliubov-Born-Green-Kirkwood-Yvon (BBGKY)
hierarchy for particle marginal densities via Spohn's formalism~\cite{spohn80}. 

This mean field limit holds at extremely low temperatures, if particles of the normal component (``thermal cloud'') of the atomic gas on average have a negligible effect on evolution. As the temperature increases, while remaining well
below the phase transition point, many particles can occupy thermally excited states. The GPWE for the condensate {\em must} be modified to account for the occupation of such states.\looseness=-1

\subsection{On past works}
\label{subsec:past}
The stationary case of finite temperatures well below the phase transition was treated systematically  by Lee and Yang in the periodic setting~\cite{leeyang58,leeyang59}. Specifically, the fraction, $\xi$, of particles at the condensate was introduced as a fixed
parameter in an approximation scheme for the $N$-body Hamiltonian; this $\xi$ was finally determined via a statistical average as a function of temperature, $T$~\cite{leeyang58}. A key idea in the approximation scheme was to treat the fluctuations about the average number, $N(1-\xi)$, of particles out of the condensate as small. This idea permeates our present treatment; see also~\cite{Margetis12}.

In the early 1980s, studies of Bosons in non-translation-invariant settings were motivated by experiments
involving spin-polarized atomic hydrogen inside magnetic traps. These studies focused on bound states. For example, by minimizing a certain temperature-dependent free-energy functional of
the self-consistent Hartree-Fock and Bogoliubov theory, as described, e.g.,
in~\cite{dorreetal}, Goldman, Silvera, and Leggett~\cite{goldmanetal} obtained equations of motion for the time-independent condensate and single-particle excitation wave
functions.  
A similar method of variational character was followed by Huse and Siggia \cite{huseetal}, who systematically derived equations of motion that retain the orthogonality between the condensate and thermally excited states. 

Another approach in stationary settings by Oliva
\cite{oliva} invoked the local-density approximation, by which
the dependence of requisite functionals on the local particle density was
assumed to be that of the homogeneous system; see also~\cite{chouyang}. Later works, e.g..~\cite{Giorgini97}, made use of Popov's diagrammatic approach~\cite{popovfadeev} for the homogeneous gas; cf.~\cite{Pitaevskii03} for a review. More recent treatments rely on PDE models that couple the condensate with the
heat bath via non-equilibrium methods of quantum kinetic theory~\cite{Zarembaetal99,Davisetal01,Baoetal04,BlakieDavis05}. For a review of various models, see, e.g.~\cite{Proukakis08}.

\subsection{Main approach and mean field evolution laws}
\label{subsec:mf-intro}
Our approach to deriving the mean field limit  
relies on: (a) a perturbative approximation scheme for the $N$-body Hamiltonian in the spirit of~\cite{leeyang58} in the non-translation-invariant framework of~\cite{Wu61} for zero temperature; and (b) a time-dependent ansatz for the $N$-body state vector that incorporates the distribution of particles over thermally excited states according to Bose statistics. 
The scheme is expected to hold when: $\delta=\sqrt{\varrho a^3}\ll 1$, where $\varrho$ is the gas local density and $a$ is the scattering length; and fluctuations about the average number of particles at each thermally excited state are small enough~\cite{leeyang58} (see section~\ref{subsec:pert} for details). The (small) parameter $\delta$ expresses the gas diluteness.

Our program is carried out in the Bosonic Fock space $\mathbb{F}$ (defined as the Hilbert space with an indefinite number of Bosons),
by making use of the fraction, $\xi$, of the particles at the condensate as a fixed 
parameter in the approximation scheme~\cite{leeyang58}; cf.~\cite{MargetisThesis}. Our treatment yields low-dimensional {\em evolution} laws; and, thus, is in principle distinct from previous treatments of bound states such as the variational approach of~\cite{goldmanetal,huseetal}, the local-density
approximation of~\cite{oliva,chouyang}, and the finite-temperature theory of~\cite{Giorgini97}.  

By starting with the $N$-body dynamics, we derive the following evolution PDEs for the condensate
wave function, $\tPh(t,x)$, and single-particle excitation wave functions, $\tph_j(t,x)$ ($j=1,\,2,\,\ldots$):
\begin{subequations}\label{eq:Phi-pdes}
\begin{align}
i\partial_t \tPh&=\{-\Delta_x+V_e(x)+g(x)[\varrho_s(t,x)+2\varrho_n(t,x)]-{\textstyle\frac{1}{2}}\xi^2\zeta(t)\}\tPh~,\label{eq:Phi-pde-Intro}\\
i\partial_t \tph_j&=\{-\Delta_x+V_e(x)+2g(x)[\varrho_s(t,x)+\varrho_n(t,x)]-{\textstyle\frac{1}{2}}\xi^2\zeta(t)\}\tph_j\nonumber\\
&\mbox{}\quad -\check b_j(t) \,\tPh(t,x)~;\label{eq:phij-pde-Intro}
\end{align}
\end{subequations}
which for bound states reduce to the corresponding laws in~\cite{huseetal}.
We assume that $\Vert \tPh(t,\cdot)\Vert_{L^2(\mathbb{R}^3)}^2=N$, $\langle \tph_i(t,\cdot),\tph_j(t,\cdot)\rangle_{L^2(\mathbb{R}^3)}=\delta_{ij}$ and $\langle \tPh(t,\cdot),\tph_j(t,\cdot)\rangle_{L^2(\mathbb{R}^3)}=0$. In the above, $\Delta_x$ denotes the 1-particle Laplacian; $V_e(x)$ is the external potential, assumed to be smooth and grow rapidly for large $|x|$; $g(x)=8\pi a(x)$ where $a(x)$ is the spatially varying scattering length; $\varrho_s$ is the mean field condensate (superfluid) density, $\xi|\tPh|^2$; $\varrho_n$ is the thermal cloud (normal fluid) density [cf.~\eqref{eq:A-def}]; $\check b_j(t)$ comes from the orthogonality of each $\tph_j$ to $\tPh$ [cf.~\eqref{eq:betaj-def}]; and 
\begin{equation}\label{eq:zeta-def}
\zeta(t)=N^{-1}\int_{\mathbb{R}^3}\dx\,g(x)\,|\tPh(t,x)|^4~.
\end{equation}
The parameter $\xi$ can in principle be determined as a function of temperature, $T$, via imposing Bose statistics for the occupation numbers of particles at thermally excited states 
(see section~\ref{sec:stat})~\cite{goldmanetal}. 
At zero temperature, when $\xi=1$ and $\rho_n\equiv 0$,~\eqref{eq:Phi-pde-Intro} readily reduces to the GPWE~\cite{Gross61,Pitaevskii61,Wu61}.

For bound states, we set  $\tPh(t,x)=e^{-iE t}\Phi(x)$ and 
$\tph_j(t,x)=e^{-iE_j t}\phi_j(x)$ where $E$ and $E_j$ are the associated energies per particle (section~\ref{sec:stat}).

\subsection{Periodic microstructure and homogenization}
\label{subsec:per}
Following~\cite{Fibich06}, we introduce the function
\begin{equation}\label{eq:per-micro}
g(x)=8\pi a(x)=g_0[1+A(x/\epsilon)]~,\qquad g_0=8\pi a_0>0~,
\end{equation}
in~\eqref{eq:Phi-pdes}; $A(y)$ is a smooth periodic function of zero mean and period equal to unity. 
If $l_{sc}$ is the length over which the scattering length varies and $l_c$ is the particle correlation length, then $\epsilon=l_{sc}/l_c$. We assume that: $l_{sc}$ is small compared to other length scales of the problem such as $l_c$, $a$, the thermal de Broglie wavelength and mean interparticle distance; and the external potential, $V_e(x)$, does not depend on the subscale $\epsilon$.

For bound states, we formally apply the two-scale expansions
\begin{equation}\label{eq:2-scale}
u(x)=u^{(0)}(\bx, x)+\sum_{k=1}^\infty\epsilon^k \, u^{(k)}(\bx,x)~,\qquad \bx=x/\epsilon~,
\end{equation}
where $u=\Phi$ or $\varphi_j$; $u^{(k)}=\mathcal O(1)$ as $\epsilon\downarrow 0$. Then, we
reduce~\eqref{eq:Phi-pdes} to a set of $\epsilon$-independent equations of motion 
for $u^{(k)}(\bx,x)$ with $k=0,\,1,\,2$ (section~\ref{sec:2-scale})~\cite{hom-book,stuart-book}. \looseness=-1

This microstructure has been inspired by a model of Fibich, Sivan and Weinstein~\cite{Fibich06}. Notably, our formal analysis pays attention to system~\eqref{eq:Phi-pdes} of defocusing nonlinear Schr\"odinger-type PDEs, to be contrasted with the single focusing PDE studied 
in~\cite{Fibich06}. 

\subsection{Limitations}
\label{subsec:limitations}
The present work points to several pending issues. Our approach is heuristic, placing emphasis on modeling aspects of the Boson gas; the rigorous analyses of the mean field limit and homogenization procedure are not pursued. We do not carry out any numerical simulation for the derived PDEs; this task is left for future work. 
In the periodic homogenization program, we focus on stationary states; the time-dependent problem, in which two time scales should be accounted for, remains unresolved. The rapid spatial variation of the scattering length, $a$, can be driven by a corresponding variation of the external potential, $V_e$, in optical traps~\cite{chin10}. The treatment of the microstructure of both $V_e$ and $a$ is the subject of near-future work.
We herein restrict attention to mean field laws, leaving out effects of quantum fluctuations due to the 
excitation of atoms in pairs from the condensate~\cite{Wu61}. The description of this process for a scattering length of periodic microstructure at zero temperature  is provided in~\cite{Margetis12}. In the present case of finite temperatures, it is expected that PDEs~\eqref{eq:Phi-pdes}  would need to be coupled with a nonlocal PDE for the requisite ``pair-excitation kernel'' which should depend on the parameter $\xi$. The modeling of this more demanding process is not touched upon in our treatment.

\subsection{Structure of article}
\label{subsec:organiz}
Section~2 presents some terminology and notation. In section~\ref{sec:background}, we review concepts of the second quantization, which are needed in  our derivations. Section~\ref{sec:mf} focuses on the derivation of mean field evolution laws. In section~\ref{sec:stat}, we focus on the formalism of stationary states. In section~\ref{sec:2-scale}, we homogenize the stationary equations of motion for a scattering length of periodic microstructure. Section~\ref{sec:discussion} concludes our article with a summary of results and open problems. \looseness=-1

\section{Notation and terminology}
\label{sec:notation}
We adopt the following conventions throughout this article, in the spirit of~\cite{Margetis12}.

\begin{itemize}

\item $\mathbb{C}$ is the complex plane, $\mathbb{Z}$ is the set of all integers, $\mathbb{N}=\{0,\,1,\,\ldots\}$, and $\mathbb{N}_*=\mathbb{N}\setminus \{0\}$. The star (${}^*$) as a superscript implies Hermitian conjugation.

\item $L_s^2(\mathbb{R}^{3n})$ is the space of symmetric (Bosonic) $L^2$ functions on $\mathbb{R}^{3n}$, which are invariant under permutations
of the particle spatial coordinates, $(x_1,\ldots, x_n)$.

\item $\langle F,G\rangle$ is the one-particle $L^2$-inner product, $\int_{\mathbb{R}^3} F(x)G(x)\,{\rm d}x$,
with induced norm $\Vert F\Vert$. The scalar product in the Fock space, $\mathbb{F}$, is denoted $\langle \cdot,\cdot\rangle_{\mathbb{F}}$ (see section \ref{subsec:Fock}); the  induced norm is $\Vert\cdot\Vert_{\mathbb{F}}$.

\item One-particle integrals with unspecified integration ranges are meant to be integrals on $\mathbb{R}^3$. Integration by parts yields vanishing boundary terms at $\infty$.

\item $\mathbb{T}^d$ denotes the $d$-dimensional unit torus (cell), where $d=3$ for our purposes.
Functions that satisfy $A(x+e_k)=A(x)$ for all $x=(x^1,\,x^2,\,x^3)\in \mathbb{R}^3$ and $k=1,\,2,\,3$, where $e_k$'s are unit Cartesian vectors, are referred to as $1$-periodic.
$\langle A\rangle$ is the cell average (or mean) of the $1$-periodic $A(x)$.

\item $H^1$ denotes the Sobolev space $W^{k,p}$ for $k=1$ and $p=2$, with dual space $H^{-1}$; and
$H^1_{\rm av}$ is the space of $H^1$ $1$-periodic functions with zero average.

\item The dual space $H^{-1}_{\rm av}(\mathbb{T}^d)=\{f\in H^{-1}(\mathbb{T}^d)\big|\,\langle f\rangle=0\}$
is the Hilbert space equipped with the inner product $\langle f,h\rangle_{H_{\rm av}^{-1}(\mathbb{T}^d)}=\langle (-\Delta)^{-1}f,h\rangle_{L^2(\mathbb{T}^d)}$.
$\Vert A \Vert_{-1}$ denotes the $H_{\rm av}^{-1}$-norm of the $1$-periodic $A(x)$; typically, $d=3$.

\item Let (the $1$-periodic) $F$ be in $L^2(\mathbb{T}^3)$. If $F$ has zero mean, then $(-\Delta_x)^{-s}F(x)=\sum_{\ell\neq 0}[\widehat{F}_\ell/(4\pi^2 |\ell|^2)^s] e^{i2\pi \ell\cdot x}$, $s>0$, where $\sum_{\ell\neq 0}\widehat{F}_\ell\,e^{i2\pi \ell\cdot x}$ is the Fourier series representing $F$. Note that $\langle (-\Delta)^{-s}F \rangle=0$, $\ell=(\ell^1,\,\ell^2,\,\ell^3)\in \mathbb{Z}^3$. Here, $\ell\cdot x$ is the (Euclidean) scalar product of the
$3$-dimensional vectors $\ell$ and $x$.

\item Writing $f=\mathcal O(g)$ [$f=o(g)$], where $f$ and $g$ are functions, means that $|f/g|$ is bounded by a nonzero constant (tends to zero) in a prescribed limit. $f\sim g$ is used loosely to imply $f-g=o(g)$. If $\mathcal A$ and $\mathcal B$ are operators on $\fc$, $\mathcal A\sim \mathcal B$ means that  $\langle \tPs, \mathcal A\tPs\rangle_{\mathbb{F}} \sim \langle \tPs, \mathcal B\tPs\rangle_{\mathbb{F}}$ where $\tPs(t)$ is any Schr\"odinger state vector with components on $n$-particle sectors of $\fc$ and $n\gg 1$.

\item $u^{(k)}$ is the $k$-th order coefficient in the two-scale expansion 
for $u=\Phi,\,\phi_j$; cf.~\eqref{eq:2-scale}. We use the symbols $f^k$ and $f_j^k$ to denote the {\em slowly-varying} parts of $\Phi^{(k)}$ and $\phi_j^{(k)}$ (section~\ref{subsec:effective}); analogous notation is used for the coefficients, $\varrho_s^{(k)}$ and $\varrho_n^{(k)}$, in the expansions for the superfluid and normal fluid densities.

\item The dot on top of a symbol denotes time derivative.

\end{itemize}

\section{Many-body Boson dynamics: Background}
\label{sec:background}
In this section, we describe the many-particle Hamiltonian (section~\ref{subsec:part}), review basics of the Fock space formalism (section~\ref{subsec:Fock}), and outline the idea of the many-body perturbation scheme (section~\ref{subsec:pert}); for related reviews, see \cite{Berezin,Solovej-summer}.

\subsection{Many-particle Hamiltonian}
\label{subsec:part}

The starting point is the Hamiltonian operator, $H_N$, of $N$ Bosons [$H_N:\ L_s^2(\mathbb{R}^{3N})\to L_s^2(\mathbb{R}^{3N})]$, viz.,
\begin{equation}
H_N=\sum_{j=1}^N [-\Delta_j+V_{\rm e}(x_j)]+\frac{1}{2}\sum_{i\neq j} \mathcal V_{int}(x_i,x_j)\qquad (x_j\in\mathbb{R}^3)~,\label{eq:H}
\end{equation}
which we will refer to as the ``PDE Hamiltonian''. The units are chosen such that $\hbar=2m=1$ ($\hbar$: Planck's constant, $m$: atomic mass) and $x_j$ are particle positions in $\mathbb{R}^3$. The interaction potential
$\mathcal V_{int}$ is positive, symmetric, and compactly supported. We assume that the external potential, $V_e(x)$, is positive, smooth and increasing rapidly for large $|x|$ with 
$V_e(x)\to \infty$ as $|x|\to\infty$.

In the following, we replace $\mathcal V_{int}$ by 
\begin{equation}
\mathcal V_{int}(x_i, x_j)\rightarrow V(x_i,x_j)= g(x_i)\ \delta(x_i-x_j)~,
\label{eq:fermi-mod}
\end{equation}
where $g(x)=8\pi a(x)>0$ and $\delta(x)$ is the Dirac mass in $\mathbb{R}^{3}$. Hence, the scattering length, $a(x)$, enters our model
as an {\em ad hoc} parameter in the spirit of~\cite{Wu61}. This is to be contrasted with the rigorous approach in~\cite{Elgart06,Erdos06,Erdos07} where the (constant) scattering length emerges from the limiting procedure.

\subsection{Fock space formalism}
\label{subsec:Fock}

The Bosonic Fock space, $\fc$, consists of vectors $\upsilon$  formed by sequences $\{\upsilon^{(n)}\}$ of $n$-particle symmetric wave functions; $\upsilon^{(n)}\in L_s^2(\mathbb{R}^{3n}$), $n\in \mathbb{N}$~\cite{Berezin}. In particular, $\Omega:=\{1,\,0,\,\ldots \}$, $\Omega\in \fc$, is the vacuum state, which
has no particles. The $N$-particle Schr\"odinger state vector on $\fc$ is represented by $\tPs(t)=\{\upsilon^{(n)}\}$
with $\upsilon^{(n)}=\tPs_N \delta_{n,N}$; $\tPs_N(\cdot,t)\in L^2_s(\mathbb{R}^{3N})$.
The scalar product on $\fc$ is defined by
$\langle\upsilon_1,\, \upsilon_2\rangle_{\mathbb{F}}=\sum_{n\ge 0}\langle\upsilon_1^{(n)}, \upsilon_2^{(n)}\rangle_{L^2(\mathbb{R}^{3n})}$.

For the one-particle wave function $f\in L^2(\mathbb{R}^3)$,
the creation and annihilation operators $a^*_f$ and $a_f$ on $\fc$ are defined by
\begin{align*}
(a^*_f\upsilon)^{(n)}(\vec{x}_n)&=n^{-1/2}\sum_{j=1}^nf(x_j)\upsilon^{(n-1)}(x_1,\ldots,x_{j-1},x_{j+1},\ldots,x_n)~,
\nonumber\\
(a_f\upsilon)^{(n)}(\vec{x}_n)&=\sqrt{n+1}\int {\rm d}x\,f^*(x)\upsilon^{(n+1)}(x,\vec{x}_n)~;
\end{align*}
$\vec{x}_n=(x_1,\ldots\,,x_n)$.
Thus, $[a_f,a^*_g]:=a_f a^*_g- a^*_g a_f=\langle f,g\rangle$ and
$[a_f, a_g]=0$. The operators $a_f^*$ and $a_f$ create and annihilate a particle at state $f$, respectively. The Boson field operator $\psi(x)$ and its adjoint, $\psi^*(x)$, are defined via
\begin{equation}
a^*_f=\int{\rm d}x\,f(x)\,\psi^*(x)~,\qquad a_f=\int {\rm d}x\,f^*(x)\,\psi(x)~;
\label{eq:psi-def}
\end{equation}
$\psi^*(x)$ ($\psi(x)$) creates (annihilates) a particle at position $x$ and is time-independent in the Schr\"odinger picture.
Note the canonical commutation relations $[\psi(x), \psi^*(y)]=\delta(x-y)$ and $[\psi(x),\psi(y)]=0$.
Evidently, $\psi(x)\Omega=0$. The (Hermitian) particle number operator,
$\mathcal N$, on $\fc$ is defined by $\mathcal N=\int{\rm d}x\,\psi^*(x)\,\psi(x)$ which obeys $(\mathcal N\upsilon)^{(n)}=n\upsilon^{(n)}$, for every $\upsilon\in \fc$ and $n\in \mathbb{N}$.

In view of this formalism, the PDE Hamiltonian, $H_N$, corresponds to
the operator $\mathcal H: \fc \,\to\, \fc$ where $(\mathcal H\upsilon)^{(n)}=H_n\upsilon^{(n)}$; specifically, by~\eqref{eq:H} and~\eqref{eq:fermi-mod},
\begin{equation}
\mathcal H=\int \dx\ \psi^*(x)[-\Delta_x+V_{\rm e}(x)]\psi(x)+\frac{1}{2}\int \dx\ \psi^*(x)^2 g(x) \psi(x)^2~.
\label{eq:H-Fock}
\end{equation}
We will henceforth use this $\mathcal H$ in the place of the PDE Hamiltonian.

\subsection{Many-body perturbation}
\label{subsec:pert}

The starting point of the perturbation scheme is to split $\psi(x)$ as~\cite{Wu61}
\begin{equation}
\psi(x)=\psi_0(t,x)+\ps1(t,x)~,
\label{eq:split}
\end{equation}
where $\psi_0$ is the Boson field annihilation operator for the condensate and 
$\ps1(t,x)$ is the Boson field annihilation operator in the space orthogonal to the condensate; $\int {\rm d}x\,\tPh^*(t,x)\ps1(t,x)=0$ where $\tPh$ is the condensate wave function. By defining $a_0(t)=N^{-1/2}a_{\tPh}$ with $\Vert\tPh\Vert_{L^2(\mathbb{R}^3)}^2=N$, we set
\begin{equation}
 \psi_0(t,x)=N^{-1/2}a_0(t)\tPh(t,x)~;
 \label{eq:psi0}
 \end{equation}
$[a_0(t), a_0^*(t)]=1$ and
$a_0(t)\Omega=0$. The operator $\ps1$ can be expanded as
\begin{equation}
\ps1(t,x)=\sum_{j=1}^\infty \tph_j(t,x)a_j(t)~,\label{eq:psi1-exp}
\end{equation}
where (slightly abusing notation) $a_j(t)$ is used in place of $a_{\tph_j}(t)$ to denote the annihilation operator for a particle at state $\tph_j$. Together with its adjoint, $a_j^*(t)$, every $a_j(t)$ satisfies the canonical commutation relations for Bosons: $[a_i^*(t), a_j(t)]=\delta_{ij}$ and $[a_i(t), a_j(t)]=0$. Hence, the operator $a_j^*(t)a_j(t)$ on Fock space represents the number of particles at state $\tph_j$.

The perturbation analysis relies on expanding $\mathcal H$ in powers of $\ps1$ and $\ps1^*$. This expansion needs to be combined with the following formal statement.
\begin{equation}
a_0^*(t)a_0(t)+\int {\rm d}x\ \ps1^*(t,x)\, \ps1(t,x)=N~,
\label{eq:N-op}
\end{equation}
which sets the particle number operator, $\mathcal N=\int{\rm d}x\,\psi^*(x)\,\psi(x)$, equal to $N$ (times the unity operator). This statement poses the constraint that the many-body Schr\"odinger state vector, $\tPs(t)$, be an eigenvector of $\mathcal N$ with eigenvalue equal to $N$ for every $t>0$;
$\mathcal N\tPs(t)=N \tPs(t)$. 
In~\eqref{eq:N-op}, $\int \dx\,\ps1^*(t,x)\,\ps1(t,x)$ is the operator for the number of particles out of the condensate. 

At this stage, it is useful to provide the following definition.
\smallskip

\begin{definition}\label{def:xi}
(Condensate fraction, $\xi$). The condensate fraction, $\xi$, is defined as the parameter equal to $\langle \tPs(t),\{a_0^*(t)a_0(t)/N\}\tPs(t)\rangle_{\mathbb{F}}$, where $\tPs(t)$ is the Schr\"odinger state vector, $\tPs(t) \in\F$; for $T\ge 0$, $0<\xi\le 1$ (see section~\ref{sec:mf}). 
\end{definition}
\smallskip

This $\xi$ will be invoked as a $T$-dependent parameter in the many-body perturbation scheme of section~\ref{sec:mf}.  By~\eqref{eq:N-op}, the average total number of particles at thermally excited states is $\langle \tPs(t),\int \dx\,\ps1^*(t,x)\,\ps1(t,x) \tPs(t)\rangle_{\F}= N(1-\xi)$. 
\smallskip

\begin{definition}\label{def:qf}(Quantum fluctuations in thermally excited states). Consider the operator-valued occupation number $\mathcal N_j(t)=a_j^*(t)a_j(t)$ ($j\in \mathbb{N}_*$) on $\F$ for each thermally excited state $\tph_j$.
The operator $\delta\mathcal N_{j}(t):=\langle \tPs(t), \mathcal N_j\tPs(t)\rangle_{\F}-\mathcal N_j$ on $\F$ describes quantum fluctuations in the number of particles at state $\tph_j$. In this vein, the operator 
$$
\delta\mathcal N_{\perp}(t):=N(1-\xi)-\int \dx\,\ps1^*(t,x)\ps1(t,x)$$ 
expresses quantum fluctuations  in the total number of particles occupying thermally excited states.
\end{definition}
\smallskip

At this point, we introduce the following assumption, which serves as a rule for our heuristics.
\smallskip

\begin{remark}\label{rmk:2}
The number of particles at each thermally excited state $\tph_j$ undergoes small fluctuations about its average value. In other words, the $\delta\mathcal N_{j}(t)$ of Definition~\ref{def:qf} is treated as ``small'', in some appropriate sense, in our approximation scheme (see section~\ref{sec:mf} for details).
\end{remark}
\smallskip

A consequence of the above assumption is the following statement.
\smallskip

\begin{remark}\label{rmk:1} 
The operator $\delta\mathcal N_\perp$ of Definition~\ref{def:qf} is treated as sufficiently ``small'' in our approximation scheme; by~\eqref{eq:N-op}, ditto for the operator $N\xi-a_0^*(t)a_0(t)$.
\end{remark}
\smallskip

At $T=0$, where $\xi\simeq 1$, the mean field limit is consistent with treating $\ps1^*\ps1$ as small, in some appropriate sense, and retaining up to linear-in-$\ps1$, $\ps1^*$ terms in $\mathcal H$ via
a tensor-product ansatz for $\tPs$~\cite{Margetis12,Wu61}. This program must be modified for nonzero temperature, $T\neq 0$: Because of the presumed presence of many, up to $\mathcal O(N)$, particles at the thermal cloud, {\em quadratic-in-$\ps1$-and-$\ps1^*$ terms must appropriately be retained in $\mathcal H$} (see section~\ref{subsec:app-H}). These terms will be collected in view of Remarks~\ref{rmk:2} and~\ref{rmk:1}.

\section{Derivation of mean field evolution laws}
\label{sec:mf}

In this section, we derive mean field laws for the condensate and one-particle thermally excited states for {\em given} $\xi$ and occupation numbers, $n_j$, of such excited states. For this purpose, we replace the many-body Hamiltonian $\mathcal H$ on $\fc$ by an operator that contains up to certain quadratic-in-$\ps1$-and-$\ps1^*$ terms, consistent with the presence of $N(1-\xi)$ particles at thermally excited states, where $0<\xi\le 1$ and $N\gg 1$ (section~\ref{subsec:app-H}). This replacement involves a sequence of steps enforcing constraint~\eqref{eq:N-op} under Remarks~\ref{rmk:2} and~\ref{rmk:1}. Evolution PDEs~\eqref{eq:Phi-pdes} in 3+1 dimensions are extracted by enforcement of the many-body Schr\"odinger  equation on an uncontrolled but self-consistent ansatz for the Schr\"odinger state vector $\tPs(t)$ (section~\ref{subsec:lowest}).

\subsection{Approximations for $N$-body Hamiltonian} 
\label{subsec:app-H}

The Fock space Hamiltonian~\eqref{eq:H-Fock} is recast to the form~\cite{MargetisThesis}
\begin{equation}\label{eq:H-I}
\mathcal H=\mathcal T + \mathcal V~,
\end{equation}
where 
\begin{equation}
\mathcal T=\int \dx\,\{|{\bf \nabla}\psi(x)|^2+V_e(x)\psi^*(x)\psi(x)\}\label{eq:T-def}
\end{equation}
and
\begin{equation}
\mathcal V=\frac{1}{2}\int \dx\,\psi^*(x)^2g(x)\psi(x)^2~.\label{eq:V-def}
\end{equation}

\subsubsection{Interaction operator, $\mathcal V$}

First, we concentrate on the interaction operator given by~\eqref{eq:V-def}.  By use of splitting~\eqref{eq:split}, $\mathcal V$ is directly decomposed as
\begin{equation}
\mathcal V=\mathcal V_0+\mathcal V_1+\mathcal V_2+\mathcal V_3+\mathcal V_4~,\label{eq:V-dec}
\end{equation}
where $\ps1^*(t,x)$ and $\ps1(t,x)$ explicitly appear $k$ times in $\mathcal V_k$ ($k=0,\,1,\,2,\,3,\,4$). Specifically, we obtain the terms
\begin{subequations}\label{eq:V-dec-expl}
\begin{eqnarray}
\mathcal V_0&=&{\textstyle\frac{1}{2}}N^{-1}\zeta(t) a^*_0(t)^2a_0(t)^2~,\label{eq:V0}\\
\mathcal V_1&=&N^{-3/2}a^*_0(t)\int \dx\,g(x)|\tPh(t,x)|^2\{\tPh^*(t,x)a^*_0(t)\ps1(t,x)
\nonumber\\
&&\mbox{} +\tPh(t,x)\ps1^*(t,x)a_0(t)\}a_0(t)~,\label{eq:V1}\\
\mathcal V_2&=&{\textstyle\frac{1}{2}}N^{-1}\int \dx\,g(x)\{\tPh^*(t,x)^2 a^*_0(t)^2\ps1(t,x)^2 +\tPh(t,x)^2\ps1^*(t,x)^2 a_0(t)^2\nonumber\\
&&\mbox{}+4|\tPh(t,x)|^2a^*_0(t)\ps1^*(t,x)\ps1(t,x)a_0(t)\},\label{eq:V2}\\
\mathcal V_3&=& N^{-1/2}\int \dx\,g(x)\{\tPh^*(t,x)a^*_0(t)\ps1^*(t,x)\ps1(t,x)^2\nonumber\\
&&\mbox{} +\tPh(t,x)\ps1^*(t,x)^2\ps1(t,x)a_0(t)\}~,\label{eq:V3}\\
\mathcal V_4&=&{\textstyle \frac{1}{2}}\int \dx\,g(x)\ps1^*(t,x)^2\ps1(t,x)^2~,\label{eq:V4}
\end{eqnarray}
\end{subequations}
where $\zeta(t)$ is defined by~\eqref{eq:zeta-def}. Equations~\eqref{eq:V-dec-expl} need to be further
manipulated for the extraction of their full dependence on $\ps1$ and $\ps1^*$. Specifically, constraint~\eqref{eq:N-op} imposes a relation between $a_0^*a_0$ and $\ps1^*\ps1$ that has not yet been accounted for in~\eqref{eq:V-dec-expl}; this task will be carried out in the sequence of steps described below.

Next, we show that all terms $\mathcal V_k$ contribute to the mean field limit, in contrast to the zero-temperature case~\cite{Margetis12}. This has been ascertained in the periodic setting at $T> 0$ in which thermally excited states correspond to nonzero momenta~\cite{leeyang58}. 

The $\mathcal V_1$ of~\eqref{eq:V1} is the simplest term to deal with since in this term we can simply set $a^*_0(t)a_0(t)= N\xi$, according to Definition~\ref{def:xi} and Remark~\ref{rmk:1}.  Thus, we make the replacement
\begin{align}
\mathcal V_1&\sim N^{-1/2}\xi\int \dx\, g(x)|\tPh(t,x)|^2
\{\tPh^*(t,x)a^*_0(t)\ps1(t,x)\nonumber\\
&\mbox{} \quad +\tPh(t,x)\ps1^*(t,x)a_0(t)\}~.\label{eq:V1-sim}
\end{align}
For $\mathcal V_0$, the procedure of approximation takes into account~\eqref{eq:N-op} and Remark~\ref{rmk:1}:
\begin{eqnarray}
\mathcal V_0&=&{\textstyle\frac{1}{2}}N^{-1}\zeta(t)\bigl\{\bigl(a_0^*(t)a_0(t)\bigr)^2-a_0^*(t)
a_0(t)\bigr\}\nonumber\\ 
&\sim& {\textstyle\frac{1}{2}}N^{-1}\zeta(t)\{a_0^*(t)a_0(t)\}^2\nonumber\\ 
&\sim&{\textstyle\frac{1}{2}}N^{-1}\zeta(t)\left\{N\xi + \left[N(1-\xi) -\int \dx\,\ps1^*(t,x)\ps1(t,x)\right]\right\}^2\nonumber\\
&\sim&{\textstyle\frac{1}{2}}N^{-1}\zeta(t)\left\{N^2\xi^2+2N\xi\left[N(1-\xi) -\int
\dx\,\ps1^*(t,x)\ps1(t,x)\right]\right\}\nonumber\\
&=&{\textstyle\frac{1}{2}}N^{-1}\zeta(t) \left[N^2\xi(2-\xi)-2N\xi\int \dx\,\ps1^*(t,x)\ps1(t,x)\right]~.\label{eq:V0-sim}
\end{eqnarray}
The term $a_0^*(t)a_0(t)$ inside the braces in the first
line of~\eqref{eq:V0-sim} contributes $\mathcal O(N\xi)$ and $\mathcal O(1)$ terms which can both be neglected in comparison to the bracketed terms of the last line. (We treat $N(1-\xi)-\int \dx\ \ps1^*\ps1$ as small compared to $N\xi$, by Remark~\ref{rmk:1}.)

Next, we address the term $\mathcal V_4$ of \eqref{eq:V4}. In view of expansion~\eqref{eq:psi1-exp}, this term is rewritten as
\begin{align}
\mathcal V_4&={\textstyle\frac{1}{2}}\sum_{i,j,k,l=1}^\infty\,a^*_i(t)a^*_j(t)a_k(t)a_l(t)\nonumber\\
&\mbox{}\quad\times\int \dx\,g(x)\tph^*_i(t,x)\tph^*_j(t,x)\tph_k(t,x)\tph_l(t,x)
\nonumber\\
&\sim \sum_{i,j=1}^{\infty}\,a^*_i(t)a^*_j(t)a_i(t)
a_j(t)\int \dx\,g(x)|\tph_i(t,x)\tph_j(t,x)|^2\nonumber\\
&\sim \int \dx\, g(x)\left[\sum_{j=1}^{\infty}\,|\tph_j(t,x)|^2 \,\mathcal N_j\right]^2~;\label{eq:V4-app}
\end{align}
recall that $\mathcal N_j(t)=a_j^*(t)a_j(t)$ is the operator for the number of particles at state $\tph_j$. The approximations in~\eqref{eq:V4-app} rely on: (i) the hypothesis that the major contribution to the sum of the first line comes from terms with $(k, l)=(i,j)$ or $(j,i)$; and (ii) the identity $a_i^*a_j^*a_ia_j=(a_i^*a_i)(a_j^*a_j)-\delta_{ij}a_i^*a_i=\mathcal N_i\mathcal N_j-\delta_{ij}a_i^*a_i$ and neglect of $\delta_{ij}a_i^*a_i$ because of its relatively small contribution. Hypothesis (i) is responsible for the appearance of the extra factor of 2 in the third line of~\eqref{eq:V4-app}. This hypothesis leads to the normal fluid density, $\varrho_n$, as an emergent effective potential in the equations of motion [see~\eqref{eq:A-def}], in the spirit of the mean field approach that we have adopted.

To further simplify $\mathcal V_4$, we apply Remark~\ref{rmk:2}. As discussed in~\cite{leeyang58}, it is desirable to start with an $N$-body state vector, $\tPs(t)$, with the occupation numbers
\begin{subequations}
\begin{equation}
n_0=\langle \tPs(t),\{a_0^*(t)a_0^*(t)\}\tPs(t)\rangle=N\xi~,\quad\quad n_j=\langle \tPs(t),\mathcal N_j(t)\tPs(t)\rangle=O(1)~,\label{eq:occ-nums}
\end{equation}
for the states $\tPh$ and $\tph_j$, respectively, where
\begin{equation}
\sum_{j=1}^{\infty}\,n_j=N(1-\xi)~;\label{eq:num-tot}
\end{equation}
\end{subequations}
and treat $\mathcal N_j-n_j$ as small compared to $n_j$ in the approximation scheme. In section~\ref{subsec:lowest}, we identify $n_j$ with occupation numbers at equilibrium.
Thus,~\eqref{eq:V4-app} becomes
\begin{align*}
\mathcal V_4&\sim \int \dx\,g(x)\left\{\sum_{j=1}^{\infty}
|\tph_j(t,x)|^2 n_j\right.\nonumber\\
&\mbox{}+\left.\left[\sum_{j=1}^{\infty}
|\tph_j(t,x)|^2\mathcal N_j(t)-\sum_{j=1}^{\infty} |\tph_j(t,x)|^2 n_j\right]\right\}^2\nonumber\\ 
&\sim \int \dx\,g(x)\left[\sum_{j=1}^{\infty}\,
|\tph_j(t,x)|^2 n_j\right] 
\left[-\sum_{j=1}^{\infty}\,|\tph_j(t,x)|^2 n_j +2\sum_{j=1}^{\infty}\, |\tph_j(t,x)|^2 \mathcal N_j(t)\right]\nonumber\\
&\sim \int \dx\,g(x)\left[\sum_{j=1}^{\infty}\,|\tph_j(t,x)|^2 n_j\right]
\left[-\sum_{j=1}^{\infty}\,|\tph_j(t,x)|^2 n_j +2\ps1^*(t,x)\ps1(t,x)\right].
\end{align*}
By~\eqref{eq:psi1-exp}, in the last line we invoked the relation
\begin{align*}
\ps1^*(t,x)\ps1(t,x)&=\sum_{i,j=1}^{\infty}\,
\tph^*_i(t,x)\tph_j(t,x)a^*_i(t)a_j(t)\nonumber\\
&\sim \sum_{j=1}^{\infty}\,|\tph_j(t,x)|^2 \mathcal N_j(t)~.\label{eq:psi1-dens}
\end{align*}

At this stage, it is natural to define the mean field density for the normal fluid as
\begin{equation}
\varrho_n(t,x)=\sum_{j=1}^{\infty}\,|\tph_j(t,x)|^2 n_j~.\label{eq:A-def}
\end{equation}
Thus, $\mathcal V_4$ is conveniently written as
\begin{equation}
\mathcal V_4\sim \int \dx\,g(x)\,\varrho_n(t,x)[-\varrho_n(t,x)+2\ps1^*(t,x)\ps1(t,x)]~.\label{eq:V4app-III}
\end{equation}

Next, we address the term $\mathcal V_2$ of~\eqref{eq:V2}. This term is split as 
\begin{equation}
\mathcal V_2=\mathcal V_{21}+\mathcal V_{22}~,\label{eq:V2-split}
\end{equation}
where $\mathcal V_{21}$ describes pair excitation, i.e., the scattering of particles in pairs from the condensate to other states and vice versa, viz.,
\begin{align}
\mathcal V_{21}&={\textstyle\frac{1}{2}}N^{-1}\int \dx\, g(x)\{\tPh^*(t,x)^2 a^*_0(t)^2\ps1(t,x)^2\nonumber\\
&\mbox{}\quad +\tPh(t,x)^2\ps1^*(t,x)^2a_0(t)^2\}~,\label{eq:V21}
\end{align}
while 
\begin{equation}
\mathcal V_{22}=2N^{-1}\int \dx\, g(x)|\tPh(t,x)|^2a^*_0(t)\ps1^*(t,x)\ps1(t,x)a_0(t)~.\label{eq:V22}
\end{equation}
In the following treatment, we neglect $\mathcal V_{21}$, since its retainment has been shown to lead to beyond-mean-field effects at $T=0$~\cite{Grillakis10,Wu61}. In contrast, $\mathcal V_{22}$ contributes to the mean field limit, and must be retained and simplified by~\eqref{eq:N-op} under Remarks~\ref{rmk:1} and~\ref{rmk:2}:
\begin{eqnarray}
\mathcal V_{22}&=&2 N^{-1}\left\{N\xi+\left[N(1-\xi) -\int \dx\, 
\ps1^*(t,x)\ps1(t,x)\right]\right\}\nonumber\\
&&\mbox{} \times \int \dx\, g(x)|\tPh(t,x)|^2\{\varrho_n(t,x)+[\ps1^*(t,x)
\ps1(t,x)-\varrho_n(t,x)]\}\nonumber\\
&\sim&2 N^{-1}\left\{N(1-\xi)\int \dx\,g(x)|\tPh(t,x)|^2 \varrho_n(t,x) \right.\nonumber\\
&&\mbox{} +N\xi\int \dx\, g(x)|\tPh(t,x)|^2\ps1^*(t,x)\ps1(t,x)\nonumber\\  
&&\mbox{} -\left.\left[\int
\dy\, g(y)|\tPh(t,y)|^2 \varrho_n(t,y)\right] \int \dx\,\ps1^*(t,x)\ps1(t,x)\right\}.\label{eq:V22-app}
\end{eqnarray}

A similar procedure can be applied to the $\mathcal V_3$ of (\ref{eq:V3}):  
\begin{eqnarray}
\mathcal V_3&=&N^{-1/2}\int \dx\, g(x)\left[\tPh^*(t,x)a^*_0(t)\sum_{i,j,k}\tph_i^*(t,x)\tph_j(t,x)\tph_k(t,x)a_i^*(t)a_j(t)a_k(t)\right.\nonumber\\
&&\mbox{} +\left. \tPh(t,x)\sum_{i,j,k}\tph_i^*(t,x)\tph_j^*(t,x)\tph_k(t,x)a_i^*(t)a_j^*(t)a_k(t)a_0(t)\right]\nonumber\\
&\sim& 2N^{-1/2}\int \dx\, g(x)\left[\tPh^*(t,x)a_0^*(t)\sum_{i,j}\tph_i^*(t,x)\tph_i(t,x)\tph_j(t,x)a_i^*(t)a_i(t)a_j(t)\right.\nonumber\\
&&\mbox{} +\left. \tPh(t,x)\sum_{i,j}
\tph_i^*(t,x)\tph_j^*(t,x)\tph_j(t,x)a_i^*(t)a_j^*(t)a_j(t)a_0(t)\right]\nonumber\\
&=&2N^{-1/2}\int \dx\,g(x)\left[\tPh^*(t,x)a^*_0(t)\,\sum_{i=1}^{\infty}\,
|\tph_i(t,x)|^2 \mathcal N_i(t)\ps1(t,x)\right.\nonumber\\
&&\mbox{} +\left.\tPh(t,x)\ps1^*(t,x)a_0(t)\,\sum_{i=1}^{\infty}\, |\tph_i(t,x)|^2 \mathcal N_i(t)\right]\nonumber\\
&\sim& 2 N^{-1/2}\int \dx\,g(x)\,\varrho_n(t,x)[\tPh^*(t,x)a^*_0(t)\ps1(t,x)\nonumber\\
&&\mbox{} \qquad +\tPh(t,x)\ps1^*(t,x)a_0(t)]~.\label{eq:V3-app}
\end{eqnarray}

Therefore, the desired approximation to the $\mathcal V$ of~\eqref{eq:V-def} is 
given by~\eqref{eq:V-dec} and~\eqref{eq:V2-split} together with
expressions \eqref{eq:V0-sim}, \eqref{eq:V1-sim}, 
\eqref{eq:V22-app}, \eqref{eq:V3-app}, \eqref{eq:V4app-III}, and $\mathcal V_{21}\simeq 0$:
\begin{align}
\mathcal V&\sim \left\{{\textstyle \frac{1}{2}}\zeta(t)N [1-(1-\xi)^2]+
2(1-\xi)\int \dx\,g(x)|\tPh(t,x)|^2 \varrho_n(t,x)\right.\nonumber\\
&\mbox{}\quad \left. -\int \dx\,g(x)\,\varrho_n(t,x)^2\right\}\nonumber\\
&\mbox{} 
+N^{-1/2}\int \dx\, g(x) \{\xi |\tPh(t,x)|^2+2\varrho_n(t,x)\}\{\tPh^*(t,x)a_0^*(t)\ps1(t,x)\nonumber\\
&\mbox{}\qquad +\tPh(t,x)\ps1^*(t,x)a_0(t)\}\nonumber\\
&\mbox{} +\int \dx\,\biggl\{-\xi\zeta(t)+2g(x)[\xi|\tPh(t,x)|^2+\varrho_n(t,x)]\nonumber\\
&\mbox{}\quad -2N^{-1}\int \dy\,g(y)|\tPh(t,y)|^2 \varrho_n(t,y)\biggr\}\ps1^*(t,x)\ps1(t,x)~.\label{eq:V-app-fin}
\end{align}

\subsubsection{Kinetic energy operator, $\mathcal T$, and resulting Hamiltonian}

The kinetic energy operator, $\mathcal T$, of~\eqref{eq:T-def} is explicitly written as
\begin{eqnarray}
\mathcal T&=&N[\zeta_\Delta(t) +\zeta_e(t)] +N^{-1/2}a_0(t)\int
\dx\, [-\Delta\tPh(t,x) +V_e(x)\tPh(t,x)]\ps1^*(t,x)\nonumber\\
&&\mbox{} +N^{-1/2}a^*_0(t)\int \dx\,[-\Delta\tPh^*(t,x) +V_e(x)\tPh^*(t,x)]\ps1(t,x)\nonumber\\
&&\mbox{} +\int \dx\,\{|{\bf \nabla}\ps1(t,x)|^2 +[V_e(x)-\zeta_\Delta(t)-\zeta_e(t)]\ps1^*(t,x)\ps1(t,x)\}~,\label{eq:T-app}
\end{eqnarray}
where $\zeta(t)$ is defined by~\eqref{eq:zeta-def} and
\begin{eqnarray}
\zeta_\Delta(t)&=&N^{-1}\int \dx\,|{\bf \nabla}\tPh(t,x)|^2~,\quad
\zeta_e(t)=N^{-1}\int \dx\,V_e(x)|\tPh(t,x)|^2~.\label{eq:zeta-delta}
\end{eqnarray}

Accordingly, the $N$-body Hamiltonian is approximated as
\begin{equation}
\mathcal H\sim \mathcal H_I=\mathcal H_0 +\mathcal H_1 + \mathcal H_2~,\label{eq:H-decomp}
\end{equation}
where
\begin{eqnarray}
\mathcal H_0&=&N[\zeta_\Delta(t) +\zeta_e(t)] +{\textstyle\frac{1}{2}}N\zeta(t)[1-(1-\xi)^2]\vphantom{\int}\nonumber\\
&&\mbox{}+2(1-\xi) \int
\dx\, g(x)|\tPh(t,x)|^2 \varrho_n(t,x) -\int \dx\,g(x)\,\varrho_n(t,x)^2\label{eq:H0-def}
\end{eqnarray}
is a $c$-number (multiplied by the unity operator), while the operators $\mathcal H_1$ and $\mathcal H_2$ depend on $\ps1$, $\ps1^*$:
\begin{eqnarray} 
\mathcal H_1&=&N^{-1/2}a_0(t)\int \dx\,\{-\Delta\tPh(t,x) +V_e(x)\tPh(t,x)\nonumber\\
&&\mbox{} +g(x)[\xi |\tPh(t,x)|^2+2\varrho_n(t,x)]\tPh(t,x)\}\ps1^*(t,x)\nonumber\\
&&\mbox{} +N^{-1/2}a^*_0(t)\int \dx\,\{-\Delta\tPh^*(t,x)+V_e(x)\tPh^*(t,x)\nonumber\\
&&\mbox{} +g(x)[\xi |\tPh(t,x)|^2+2\varrho_n(t,x)]\tPh^*(t,x)\}
\ps1(t,x)~,\label{eq:H1-app}
\end{eqnarray}
\begin{eqnarray}
\mathcal H_2&=&\int \dx\,\left\{|{\bf \nabla}\ps1(t,x)|^2 +\left[
V_e(x)-\zeta_\Delta(t)-\zeta_e(t)-\xi\zeta(t)\vphantom{\int}\right.\right.\nonumber\\
&&\mbox{}+2g(x)[\xi|\tPh(t,x)|^2+\varrho_n(t,x)]\nonumber\\
&&\mbox{}-\left.\left. 2 N^{-1}\int \dy\,g(y)|\tPh(t,y)|^2 \varrho_n(t,y)\right]\ps1^*(t,x)\ps1(t,x)\right\}~.\label{eq:H2-app}
\end{eqnarray} 
We view the resulting $\mathcal H_I$ as the ``lowest-order perturbation'' for Hamiltonian $\mathcal H$ in the present context of finite temperatures.

\subsection{Mean field PDEs}
\label{subsec:lowest}

Next, we combine~\eqref{eq:H-decomp}--\eqref{eq:H2-app} with an ansatz for the Schr\"odinger state vector $\tPs(t)$ in order to derive PDEs~\eqref{eq:Phi-pdes}. For zero temperature, $T=0$, the mean-field approximation for $\tPs(t)$ is the tensor product $\tPs(t)\simeq (N!)^{-1/2}a_0^*(t)^{N}\Omega$, in which all particles occupy the one-particle state $\tPh$. For the present case with finite temperatures, we modify this ansatz to
\begin{equation}
\tPs(t)\sim e^{-i\theta(t)}\,\frac{a^*_0(t)^{N\xi}}
{[(N\xi)!]^{1/2}}\,\prod_{j=1}^{\infty}\,\frac{a^*_j(t)^{n_j}}
{(n_j!)^{1/2}}\,\Omega~,\label{eq:Psi-tp}
\end{equation}
where $\theta(t)$ is a global phase to be determined and $\{n_j\}_{j=1}^\infty$ amounts to the occupation numbers of the thermally excited states; $a_0(t)$ and $a_j(t)$ are described in~\eqref{eq:psi0} and~\eqref{eq:psi1-exp}. We consider ansatz~\eqref{eq:Psi-tp} as a mean field approximation consistent with the lowest-order $\mathcal H_I$ of~\eqref{eq:H-decomp}--\eqref{eq:H2-app}, for small fluctuations of the $j$-th state occupation numbers, $\mathcal N_j$, about their averages. The enforcement of this consistency through the many-body dynamics of the Schr\"odinger equation should yield the desired evolution laws for $\tPh$ and $\tph_j$. 

Several comments on~\eqref{eq:Psi-tp} are in order.
\smallskip

(i) Recall that $n_j$ satisfy 
\eqref{eq:num-tot} and $n_j\in \mathbb{N}$. Therefore, only a finite number of these $n_j$ can be
nonzero. Thus, product~\eqref{eq:Psi-tp} is finite.

(ii) Since the $n_j$ are integers, $\,d n_j/dt$ can only be delta
functions.   However, no such delta
functions are present in the Hamiltonian.  Hence, in the present scheme all $n_j$ must be
time-independent. We henceforth set $n_j=\en_j$ to emphasize the equilibrium-like character of these occupation numbers.

(iii) This time independence of $\en_j$  is related to a point discussed in \cite{Wu61} for zero temperature,  namely, that the underlying approach is valid only at a moderate
time scale.  This restriction holds for nonzero temperatures as well. The issue of shorter and longer time
scales remains open.

(iv) The factor $e^{-i\theta(t)}$ is introduced for later algebraic convenience. By suitable choice of the phase, $\theta(t)$, the equations of motion
for $\tPh(t,x)$ and $\tph_j(t,x)$ can be relatively simplified.  
\smallskip

To this lowest order of perturbation theory, by neglect of pair excitation, the many-particle Schr\"odinger equation reads
\begin{equation}
i\partial_t\tPs(t)=\mathcal H_I\tPs(t)~,\label{eq:N-SE}
\end{equation}
where $\mathcal H_I$ is approximation~\eqref{eq:H-decomp} for the Fock space Hamiltonian, $\mathcal H$.
To facilitate the operator algebra implied by~\eqref{eq:N-SE}, define the following state vectors.
\begin{subequations}\label{eq:Psi-ind} 
\begin{equation}
\tPs_{(0)}(t):=e^{-i\theta(t)}N\xi\,\frac{a^*_0(t)^{N\xi-1}}{[(N\xi)!]^{1/2}}\,
\prod_{j=1}^{\infty}\,\frac{a^*_j(t)^{\en_j}}{(\en_j!)^{1/2}}\,\Omega\label{eq:Psi0}
\end{equation}
and, for $i\in \mathbb{N}_*$,
\begin{equation}
\tPs_{(i)}(t):=e^{-i\theta(t)}\,\frac{a^*_0(t)^{N\xi}}
{[(N\xi)!]^{1/2}}\,n_i^0\,\frac{a^*_i(t)^{n^{\rm eq}_i-1}}
{(n_i^0!)^{1/2}}\prod_{j\ne i}\frac{a^*_j(t)^{\en_j}}{(\en_j!)^{1/2}}\,\Omega~.\label{eq:Psi-i}
\end{equation}
\end{subequations}
Clearly, $\tPs_{(i)}(t)=0$ if $n_i^{\rm eq}\equiv 0$.  Thus, there is only a finite number
of nonzero $\tPs_{(i)}(t)$ by this formalism.  
The left-hand side of~\eqref{eq:N-SE}  equals
\begin{equation*}
i\partial_t\tPs(t)=\dot\theta(t)\tPs(t)+i\,\dot a_0^*(t)\,\tPs_{(0)}(t) +\sum_{j=1}^{\infty}\,i\,\dot a^*_j(t)\,\Psi_{(j)}(t)~.
\end{equation*}

By virtue of~\eqref{eq:H0-def}--\eqref{eq:H2-app} and after some algebra, \eqref{eq:N-SE}  reads
\begin{equation}
\dot\theta(t)\tPs(t)+\mathcal B^*_0(t)\tPs_{(0)}(t)+\sum_{j=1}^{\infty}\,\mathcal B^*_j(t)\tPs_{(j)}(t)=\mathcal H_0\tPs(t)~,\label{eq:t-SE}
\end{equation}
where
\begin{eqnarray}
\mathcal B^*_0(t)&=&i\dot a^*_0(t)-N^{-1/2}\int
\dx\, \big\{-\Delta\tPh(t,x)+V_e(x)\tPh(t,x)\nonumber\\
&&\mbox{} +\xi g(x) |\tPh(t,x)|^2\tPh(t,x) +2g(x)\varrho_n(t,x)\tPh(t,x)\big\}\ps1^*(t,x)\label{eq:B0-def}
\end{eqnarray}
and, for $j\in\mathbb{N}_*$, 
\begin{eqnarray}
\mathcal B^*_j(t)&=&i\dot a^*_j(t)-N^{-1/2}a^*_0(t)\int \dx\,\big\{-\Delta\tPh^*(t,x)+V_e(x)\tPh^*(t,x)\nonumber\\
&&\mbox{}+\xi g(x)|\tPh(t,x)|^2\tPh^*(t,x) 
+2g(x)\varrho_n(t,x)\tPh^*(t,x)\big\}\tph_j(t,x)\nonumber\\
&&\mbox{}-\int \dx\,\left\{-\Delta\tph_j(t,x)
+\left[V_e(x)-
\zeta_\Delta(t)-\zeta_e(t)\vphantom{\int}\right.\right.\nonumber\\ &&\mbox{}
-\xi\zeta(t)+2\xi g(x)|\tPh(t,x)|^2+2g(x)\varrho_n(t,x)\nonumber\\
&&\mbox{}-\left.\left. 2N^{-1}\int \dy\, g(y)|\tPh(t,y)|^2 \varrho_n(t,y)\right]\tph_j(t,y)\right\}\ps1^*(t,x)~.\label{eq:Bj-def}
\end{eqnarray}

By\ \eqref{eq:t-SE}--\eqref{eq:Bj-def}, the operators $\mathcal B_j(t)$ 
should satisfy
\begin{equation} 
\mathcal B^*_0(t)=\bar c_0(t)a^*_0(t)~,\quad
\mathcal B^*_j(t)=\bar c_j(t)a^*_j(t)~,\label{eq:motion-Bj}
\end{equation}
where $\bar c_0(t)$ and $\bar c_j(t)$ are such that
\begin{equation}
\dot \theta(t)+N\xi\bar c_0(t)+\sum_{j=1}^{\infty}\,\en_j\bar
c_j(t)=\mathcal H_0~.\label{eq:cj}
\end{equation}
Equation~\eqref{eq:cj} is the only condition that
$\dot \theta(t)$, $\,\bar c_0(t)$, and $\bar c_j(t)$ must satisfy.

Equations~\eqref{eq:motion-Bj} with~\eqref{eq:cj} yield PDEs for $\tPh(t,x)$ and $\tph_j(t,x)$, in view of~\eqref{eq:B0-def} and~\eqref{eq:Bj-def}.
Indeed, taking into account that $a^*_0(t)=N^{-1/2}\int \dx\,\tPh(t,x)\psi^*(x)$ and $\ps1^*(t,x)=\psi^*(x)-N^{-1/2}a_0^*(t)\tPh^*(t,x)$, we write relation~\eqref{eq:motion-Bj} for $\mathcal B_0^*(t)$ as
\begin{eqnarray*}
\lefteqn{\int \dx\,\psi^*(x)\Biggl\{-i\partial_t\tPh(t,x) +\bar c_0(t)\tPh(t,x)\vphantom{\int}}\quad\nonumber\\ 
&&\mbox{}+\int \dy\,[\delta(x-y) -N^{-1}
\tPh(t,x)\tPh^*(t,y)]\nonumber\\
&&\qquad \times \big[-\Delta_y+V_e(y)+\xi g(y) |\tPh(t,y)|^2 
+2 g(y) \varrho_n(t,y)\big]\tPh(t,y)\Biggr\}=0~.
\end{eqnarray*}
Thus, the quantity in the braces of the last expression must vanish, leading to
\begin{subequations}\label{eq:Phis-PDE-I}
\begin{eqnarray}
i\partial_t \tPh(t,x)
&=&\left\{-\Delta_x+V_e(x)+g(x) \xi |\tPh(t,x)|^2 
+2 g(x) \varrho_n(t,x)\right.\nonumber\\
&&\mbox{} -\left[-\bar c_0(t)+\zeta_\Delta(t)+\zeta_e(t)+\xi \zeta(t)\vphantom{\int}\right.\nonumber\\  
&&\mbox{} +\left. \left.2 N^{-1}\int \dy\,g(y)\,\varrho_n(t,y)|\tPh(t,y)|^2\right]\right\}\tPh(t,x)~.\label{eq:Phi-PDE-I}
\end{eqnarray}
In the same vein, by\ \eqref{eq:Bj-def} and~\eqref{eq:motion-Bj} for $B_j^*(t)$ ($j\in \mathbb{N}_*$), we obtain the PDE 
\begin{eqnarray}
i\partial_t\tph_j(t,x)
&=&\left[-\Delta_x+V_e(x)+2g(x) \xi |\tPh(t,x)|^2 +2g(x) \varrho_n(t,x)\vphantom{\int}\right.\nonumber\\ 
&&\mbox{} -\zeta_\Delta(t)-\zeta_e(t)- \xi \zeta(t)+\bar c_j(t)\nonumber\\ 
&&\mbox{}\left. -2 N^{-1}\int \dy\, g(y)\,\varrho_n(t,y)|\tPh(t,y)|^2\right]\tph_j(t,x) \nonumber\\
&&\mbox{}-N^{-1}\Phi(t,x)\int \dy\,
\tPh^*(t,y) g(y)\, \xi |\Phi(t,y)|^2\tph_j(t,y)~,\label{eq:phij-PDE-I}
\end{eqnarray} 
\end{subequations}
where the last term in~\eqref{eq:phij-PDE-I} results from the orthogonality of $\tPh$ and $\tph_j$. 

Equations~\eqref{eq:Phis-PDE-I} can be further simplified.  
By~\eqref{eq:cj}, we write
\begin{eqnarray*}
\lefteqn{\dot\theta(t)+N\xi \bar c_0(t)+\sum_{j=1}^{\infty}\,\en_j\bar
c_j(t)}\quad\nonumber\\
&=&N[\zeta_\Delta(t)+\zeta_e(t)]+{\textstyle \frac{1}{2}}N\zeta(t)[1-(1-\xi)^2]\nonumber\\
&&\mbox{} +2 (1-\xi)\int \dx\,g(x)|\Phi(t,x)|^2
\varrho_n(t,x) -\int \dx\,g(x) \varrho_n(t,x)^2~.
\end{eqnarray*}
By comparing the last equation to PDEs\ \eqref{eq:Phis-PDE-I},
we introduce $c_0(t)$ and $c_j(t)$ through
\begin{eqnarray*}
\bar c_0(t) -c_0(t)&=&\bar c_j(t)-c_j(t)\nonumber\\
&=&\zeta_\Delta(t) +\zeta_e(t) +{\textstyle\frac{1}{2}}\zeta(t)[1-(1-\xi)^2]\nonumber\\
&&\mbox{} +2N^{-1}\int
\dx\, g(x)|\tPh(t,x)|^2\,\varrho_n(t,x)~,\quad j\in\mathbb{N}_*~.
\end{eqnarray*}
Thus, $c_0(t)$ and $c_j(t)$ satisfy
\begin{eqnarray}
\lefteqn{\dot\theta(t)+N\xi
c_0(t)+\sum_{j=1}^{\infty}\,\en_j c_j(t)}\quad\nonumber\\
&=&-2\xi \int \dx\,g(x)|\tPh(t,x)|^2 \varrho_n(t,x)
-\int \dx\, g(x)\,\varrho_n(t,x)^2~.\label{eq:theta-c}
\end{eqnarray}
Accordingly, the governing equations of motion are recast into 
\begin{subequations}\label{eq:Phis-PDE-II}
\begin{eqnarray}
i\partial_t\tPh(t,x)&=&\left[-\Delta+V_e(x)+g(x) \,\xi |\tPh(t,x)|^2\right.
\nonumber\\
&&\mbox{} \left.+2 g(x) \varrho_n(t,x)-{\textstyle\frac{1}{2}}\xi^2\zeta(t) +c_0(t)\right]\tPh(t,x)~,\label{eq:Phi-PDE-II} 
\end{eqnarray}
\begin{eqnarray}
i\partial_t\tph_j(t,x)&=&\left[-\Delta+V_e(x) +2 g(x)\, \xi |\tPh(t,x)|^2\right.\nonumber\\
&&\mbox{}\left.+2g(x) \varrho_n(t,x)-{\textstyle\frac{1}{2}} \xi^2\zeta(t)+c_j(t)\right]\tph_j(t,x)\nonumber\\
&&\mbox{}  -N^{-1}\tPh(t,x)\int \dy\, \tPh^*(t,y) g(y)\,\xi|\tPh(t,y)|^2\tph_j(t,y)~.\label{eq:phij-PDE-II}
\end{eqnarray}
\end{subequations}

Equations\ \eqref{eq:theta-c} and~\eqref{eq:Phis-PDE-II} are the sole
consequences of the many-body Schr\"o\-dinger equation 
with a state vector of form \eqref{eq:Psi-tp}.  However, physically it does not make sense to have equations of motion for $\tph_j(t,x)$ with $c_j(t)$ depending on $j$;
each $\tph_j(t,x)$ should see the same effective potential.  Thus, $\theta(t)$ is chosen so that
\begin{displaymath}
\mbox{each}\ c_j(t)\ \mbox{is\ independent\ of}\ j~.
\end{displaymath}
We make the simplest choice, $c_j(t)\equiv 0$ for all $j\in\mathbb{N}$.
Hence, the global phase $\theta(t)$ entering~\eqref{eq:Psi-tp} obeys
\begin{equation}\label{eq:theta-prime-fin}
\dot\theta(t)=-2 \xi \int \dx\, g(x)|\tPh(t,x)|^2 \varrho_n(t,x) -\int \dx\, g(x)
\varrho_n(t,x)^2~.
\end{equation}
By~\eqref{eq:Phis-PDE-II}, we obtain PDEs~\eqref{eq:Phi-pdes} for $\tPh(t,x)$ and $\tph_j(t,x)$ 
with $\varrho_s(t,x)=\xi |\tPh(t,x)|^2$,
\begin{equation}
\check b_j(t)=N^{-1}\int \dy\, \tPh^*(t,y) g(y)\,\xi|\tPh(t,y)|^2\tph_j(t,y)~,\label{eq:betaj-def}
\end{equation}
and the thermal cloud density, $\varrho_n$, furnished by~\eqref{eq:A-def}. For {\em fixed} $\xi$ and $\{\en_j\}_{j=1}^\infty$, this description concludes our derivation. In section~\ref{sec:stat}, we propose a closure of the entire PDE system via the (equilibrium) Bose-Einstein distribution for $n_j^{\rm eq}$.

\section{Stationary states}
\label{sec:stat}

In this section, we simplify PDEs~\eqref{eq:Phi-pdes} when the particle system occupies stationary states, where $\tPh$ corresponds to the lowest macroscopic state.  
Furthermore, we apply the Bose-Einstein distribution for the occupation numbers $\en_j$~\cite{goldmanetal}. 

We set
\begin{align}
\tPh(t,x)&=\Phi(x)\, e^{-iE\,t}~,\nonumber\\
\tph_j(t,x)&=\phi_j(x)\, e^{-iE_jt}~,\label{eq:stat-solns} 
\end{align}
where $E$ is the (lowest) energy per particle of the condensate and $E_j$ is the energy per particle of the $j$th thermally excited state; $\Phi$ is assumed to be real.
By defining
\begin{align}\label{eq:mus-def}
\mu=E+{\textstyle\frac{1}{2}}\xi^2\zeta~,\quad \mu_j=E_j+{\textstyle\frac{1}{2}}\xi^2\zeta~,
\end{align}
we re-write PDEs~\eqref{eq:Phi-pdes} as 
\begin{subequations}\label{eq:phis-stat}
\begin{align}
\{-\Delta_x+V_e(x)+g(x)[\varrho_s(x)+2\varrho_n(x)])\}\Phi(x)&=\mu \Phi(x)~,\label{eq:Phi-stat}\\
\{-\Delta_x+V_e(x)+2g(x)[\varrho_s(x)+\varrho_n(x)]\}\phi_j(x)-b_j\,\Phi(x)&=\mu_j \phi_j(x)~;\label{eq:phij-stat}
\end{align}
\end{subequations}
\begin{equation}\label{eq:bj-stat}
b_j=b_j[\phi_j;\Phi]=N^{-1}\xi\int \dy\ g(y)\,\Phi(y)^3\,\phi_j(y)~;
\end{equation}
$\varrho_s=\xi\Phi^2$ and $\varrho_n=\sum_j |\phi_j|^2 \en_j$ denote the superfluid and normal fluid densities, and $\zeta$ is the static version of~\eqref{eq:zeta-def}. Recall that $\Vert \Phi\Vert^2=N$ and $\Vert \phi_j \Vert=1$.

Evidently, $\mu$ and $\mu_j$ are real. Each wavefunction $\phi_j$ can be chosen to be real. This property follows heuristically by treating $\Phi$, and hence $\varrho_n$, as fixed in~\eqref{eq:phij-stat}: Suppose $\phi_j$ is complex. For  given functions $\Phi$, $\varrho_s$ and $\varrho_n$, \eqref{eq:phij-stat} is satisfied by $c(\phi_j+\phi_j^*)$ where $c$ is any constant and $b_j[\phi_j;\Phi]$ is replaced by $b_j=b_j[c(\phi_j+\phi_j^*);\Phi]$.

By ansatz~\eqref{eq:Psi-tp} and formula~\eqref{eq:theta-prime-fin} for the global phase, $\theta(t)$, the total energy of the Boson system equals
\begin{align}\label{eq:total-en}
\mathcal E&=\dot \theta(t)+\sum_{j=0}^\infty n_j^{\rm eq} E_j\qquad (n_0^{\rm eq}=N\xi,\,E_0=E)\nonumber\\
&=N\xi E+\sum_{j=1}^\infty n_j^{\rm eq} E_j-2 \xi \int \dx\, g(x)\Phi(x)^2 \varrho_n(x) -\int \dx\, g(x) \varrho_n(x)^2~.
\end{align}

It is plausible to apply the Bose-Einstein distribution,  setting~\cite{goldmanetal}
\begin{equation}\label{eq:Bose-dis}
\en_j=\bigl( \bar z^{-1}e^{\beta \mu_j}-1\bigr)^{-1}~,\qquad j\in\mathbb{N}~,
\end{equation}
where $n_0^{\rm eq}=N\xi$, $\beta=1/(k_BT)$ ($k_BT$: Boltzmann energy) and $\bar z e^{-(1/2)\beta\,\xi^2\zeta}$ is the fugacity of the atomic gas, a Lagrange multiplier (usually denoted by  $z$)~\cite{Huang-book}. To ease notation, we remove the bar and write $z$ instead of $\bar z$.
We impose the constraints~\cite{goldmanetal}
\begin{equation}\label{eq:constr}
N\xi = \bigl( z^{-1}e^{\beta \mu}-1\bigr)^{-1}~,\quad N(1-\xi)=\sum_{j=1}^\infty \bigl( z^{-1}e^{\beta \mu_j}-1\bigr)^{-1}~.
\end{equation}
Note that by~\eqref{eq:phis-stat} $\mu$ and $\mu_j$ can in principle be determined as a function of $(\xi, z)$. Hence,~\eqref{eq:constr} should  yield each of $\xi$ and $z$ as a function of $T$.

\section{Periodic microstructure: Two-scale analysis}
\label{sec:2-scale}

In this section, we restrict attention to a scattering length that varies periodically according to~\eqref{eq:per-micro}. Recall that the fast-varying $A$ is $1$-periodic. We formally apply two-scale expansion~\eqref{eq:2-scale} to PDE system~\eqref{eq:phis-stat} with~\eqref{eq:constr} by recourse to periodic homogenization~\cite{hom-book}. The 
spatial gradient in the governing PDEs is replaced by $\nabla_x+\epsilon^{-1}\nabla_{\bx}$ ($\bx=x/\epsilon$).
 We also expand
\begin{equation*}
\varpi=\varpi^{(0)}+\sum_{k=1}^\infty \epsilon^k\, \varpi^{(k)}\qquad \mbox{as}\ \epsilon\downarrow 0~;\ \varpi=\mu,\,\mu_j,\,\xi,\,z,\,n_j^{\rm eq},\,b_j~.
\end{equation*}
A two-scale expansion ensues for the densities $\varrho_s$ and $\varrho_n$.

In our analysis, we make use of the following standard result~\cite{hom-book,Fibich06}.

\begin{lemma}\label{lem:Fred}
 Equation $-\Delta u=S(\cdot, x)$, where $S(\cdot, x)$ is $1$-periodic, admits a $1$-periodic solution $u(\cdot,x)$ only if
\begin{subequations}\label{eq:solvb}
\begin{equation}\label{eq:solvb-I}
\langle S(\cdot,x)\rangle=\int_{\mathbb{T}^d}S(\bx,x)\ \dbx=0~.
\end{equation}
The $1$-periodic solution reads as
\begin{equation}\label{eq:solvb-II}
u(\bx,x)=(-\Delta_{\bx})^{-1}S(\bx,x)+c(x)~,
\end{equation}
where $c(x)$ is reasonably arbitrary.
\end{subequations}
\end{lemma}

This lemma, stated here without proof, is a consequence of the Fredholm alternative~\cite{hom-book}.
Any solution of form~\eqref{eq:solvb-II} will be referred to as ``admissible''~\cite{Margetis12}.

\subsection{Cascade of equations}
\label{subsec:casc}

Next, we describe the set of equations for coefficients of two-scale expansion~\eqref{eq:2-scale}.
The substitution of expansion~\eqref{eq:2-scale} into PDEs~\eqref{eq:phis-stat} yields a cascade of equations for $\Phk$ and $\phjk$, which we write here for $k\le 4$:
\begin{subequations}\label{eq:casc-Phi}
\begin{align}
\mathcal O(\epsilon^{-2}):\quad &-\Delta_{\bx}\Phi^{(0)}=0=:S^{(0)}_{\Phi}~,\label{eq:casc-Phi0}\\
\mathcal O(\epsilon^{-1}):\quad &-\Delta_{\bx}\Phi^{(1)}=2\nabla_x\cdot\nabla_{\bx}\Phi^{(0)}=:S^{(1)}_{\Phi}~,\label{eq:casc-Phi1}\\
\mathcal O(\epsilon^0):\quad &-\Delta_{\bx}\Phi^{(2)}=2\nabla_x\cdot\nabla_{\bx}\Phi^{(1)}-\{-\Delta_x+V_e(x)\nonumber\\
& \mbox{}\quad +g_0[1+A(\bx)](\varrho_s^{(0)}+2\varrho_n^{(0)})-\mu^{(0)}\}\Phi^{(0)}=:S^{(2)}_{\Phi}~,\label{eq:casc-Phi2}\\
\mathcal O(\epsilon):\quad & -\Delta_{\bx}\Phi^{(3)}=2\nabla_x\cdot\nabla_{\bx}\Phi^{(2)}-\{-\Delta_x+V_e(x)+g_0[1+A(\bx)]\nonumber\\
&\mbox{}\quad \times (\varrho_s^{(0)}+2\varrho_n^{(0)})-\mu^{(0)}\}\Phi^{(1)}
\nonumber\\
&\mbox{}\quad -\{g_0[1+A(\bx)](\varrho_s^{(1)}+2\varrho_n^{(1)})-\mu^{(1)}\}\Phi^{(0)}=:S^{(3)}_{\Phi}~,\label{eq:casc-Phi3}\\
\mathcal O(\epsilon^2):\quad & -\Delta_{\bx}\Phi^{(4)}=2\nabla_x\cdot\nabla_{\bx}\Phi^{(3)}-\{-\Delta_x+V_e(x)+g_0[1+A(\bx)]\nonumber\\
&\mbox{}\quad\times (\varrho_s^{(0)}+2\varrho_n^{(0)})-\mu^{(0)}\}\Phi^{(2)}\nonumber\\
&\mbox{}\quad -\{g_0[1+A(\bx)](\varrho_s^{(1)}+2\varrho_n^{(1)})-\mu^{(1)}\}\Phi^{(1)}\nonumber\\
&\mbox{}\quad -\{g_0 [1+A(\bx)](\varrho_s^{(2)}+2\varrho_n^{(2)})-\mu^{(2)}\}\Phi^{(0)}=:S^{(4)}_{\Phi}~;\label{eq:casc-Phi4}
\end{align}
\end{subequations}
and, for $j\in \mathbb{N}_*$,
\begin{subequations}\label{eq:casc-phj}
\begin{align}
\mathcal O(\epsilon^{-2}):\quad &-\Delta_{\bx}\phi_j^{(0)}=0=:S^{(0)}_{\phi,j}~,\label{eq:casc-phj0}\\
\mathcal O(\epsilon^{-1)}:\quad &-\Delta_{\bx}\phi_j^{(1)}=2\nabla_x\cdot\nabla_{\bx}\phi_j^{(0)}=:S^{(1)}_{\phi,j}~,\label{eq:casc-phj1}\\
\mathcal O(\epsilon^0):\quad &-\Delta_{\bx}\phi_j^{(2)}=b_j^{(0)}\Phi^{(0)}+2\nabla_x\cdot\nabla_{\bx}\phi_j^{(1)}-\{-\Delta_x+V_e(x)\nonumber\\
& \mbox{}\quad +2g_0[1+A(\bx)](\varrho_s^{(0)}+\varrho_n^{(0)})-\mu_j^{(0)}\}\phi_j^{(0)}=:S^{(2)}_{\phi,j}~,\label{eq:casc-phj2}\\
\mathcal O(\epsilon):\quad &-\Delta_{\bx}\phi_j^{(3)}=b_j^{(0)}\Phi^{(1)}+b_j^{(1)}\Phi^{(0)}+2\nabla_x\cdot\nabla_{\bx}\phi_j^{(2)}-\{-\Delta_x+V_e(x)\nonumber\\
&\mbox{}\quad +2g_0[1+A(\bx)](\varrho_s^{(0)}+\varrho_n^{(0)})-\mu_j^{(0)}\}\phi_j^{(1)}\nonumber\\
&\mbox{}\quad -\{2g_0[1+A(\bx)](\varrho_s^{(1)}+2\varrho_n^{(1)})-\mu_j^{(1)}\}\phi_j^{(0)}=:S_{\phi,j}^{(3)}~,\label{eq:casc-phj3}\\
\mathcal O(\epsilon^2):\quad &-\Delta_{\bx}\phi_j^{(4)}=b_j^{(0)}\Phi^{(2)}+b_j^{(1)}\Phi^{(1)}+b_j^{(2)}\Phi^{(0)}+2\nabla_x\cdot\nabla_{\bx}\phi_j^{(3)}\nonumber\\
&\mbox{}\quad -\{-\Delta_x+V_e(x)+2g_0[1+A(\bx)](\varrho_s^{(0)}+\varrho_n^{(0)})-\mu_j^{(0)}\}\phi_j^{(2)}\nonumber\\
&\mbox{}\quad -\{2g_0[1+A(\bx)](\varrho_s^{(1)}+\varrho_n^{(1)})-\mu_j^{(1)}\}\phi_j^{(1)}\nonumber\\
&\mbox{}\quad -\{2g_0 [1+A(\bx)](\varrho_s^{(2)}+\varrho_n^{(2)})-\mu_j^{(2)}\}\phi_j^{(0)}=:S^{(4)}_{\phi,j}~.\label{eq:casc-phj4}
\end{align}
\end{subequations}

In the above, $\varrho_s^{(k)}(\bx,x)$ and $\varrho_n^{(k)}(\bx,x)$ are the $k$-th order coefficients in the expansion for the superfluid and normal fluid densities, viz.,
\begin{subequations}\label{eq:s-dens-k}
\begin{align}
\varrho_s^{(0)}(\bx,x)&=\xi^{(0)}\Phi^{(0)}(\bx,x)^2~,\label{eq:s-dens-k-a}\\
\varrho_s^{(1)}(\bx,x)&=\xi^{(1)}\Phi^{(0)}(\bx,x)^2+2\xi^{(0)}\Phi^{(0)}(\bx,x)\Phi^{(1)}(\bx,x)~,\label{eq:s-dens-k-b}\\
\varrho_s^{(2)}&=\xi^{(2)}{\Phi^{(0)}}^2+2\xi^{(1)}\Phi^{(1)}\Phi^{(0)}+\xi^{(0)}\bigl(2\Phi^{(2)}\Phi^{(0)}+{\Phi^{(1)}}^2\bigr)~;\label{eq:s-dens-k-c}
\end{align}
\end{subequations}
\begin{subequations}\label{eq:n-dens-k}
\begin{align}
\varrho_n^{(0)}(\bx,x)&=\sum_{j=1}^\infty n_j^{(0)}\,{\phi_j^{(0)}}(\bx,x)^2~,\label{eq:n-dens-k-a}\\
\varrho_n^{(1)}(\bx,x)&=\sum_{j=1}^\infty\bigl[ n_j^{(1)}\,{\phi_j^{(0)}}(\bx,x)^2 +2n_j^{(0)}\phi_j^{(0)}(\bx,x)\phi_j^{(1)}(\bx,x)\bigr]~,\label{eq:n-dens-k-b}\\
\varrho_n^{(2)}&=\sum_{j=1}^\infty\bigl\{n_j^{(2)}\,{\phi_j^{(0)}}^2 +2n_j^{(1)}\,\phi_j^{(1)}\phi_j^{(0)} +n_j^{(0)}\bigl[2\phi_j^{(2)}\phi_j^{(0)}+{\phi_j^{(1)}}^2\bigr]\bigr\}~,\label{eq:n-dens-k-c}
\end{align}
\end{subequations}
where $n_j^{(k)}$ are expansion coefficients for Bose-Einstein distribution~\eqref{eq:Bose-dis}:
\begin{subequations}\label{eq:bose-k}
\begin{align}
n_j^{(0)}&=\bigl[{z^{(0)}}^{-1}e^{\beta \mu_j^{(0)}}-1\bigr]^{-1}~,\label{eq:bose-k-a}\\
n_j^{(1)}&=\frac{{z^{(0)}}^{-1}e^{\beta\mu_j^{(0)}}}{\bigl[{z^{(0)}}^{-1}e^{\beta\mu_j^{(0)}}-1\bigr]^2}\biggl(\frac{z^{(1)}}{z^{(0)}}-\beta\mu_j^{(1)}\biggr)~,\label{eq:bose-k-b}\\
n_j^{(2)}&=\frac{{z^{(0)}}^{-1}e^{\beta\mu_j^{(0)}}}{\bigl[{z^{(0)}}^{-1}e^{\beta\mu_j^{(0)}}-1\bigr]^2}\biggl[-\biggl(\frac{{z^{(1)}}^2}{{z^{(0)}}^2}-\frac{z^{(2)}}{z^{(0)}}-\beta\mu_j^{(1)}\frac{z^{(1)}}{z^{(0)}}+\beta\mu_j^{(2)}\nonumber\\
&\mbox{}\quad +{\textstyle \frac{1}{2}}\beta^2{\mu_j^{(1)}}^2\biggr)+\frac{{z^{(0)}}^{-1}e^{\beta\mu_j^{(0)}}}{{z^{(0)}}^{-1}e^{\beta\mu_j^{(0)}}-1}\biggl(\frac{z^{(1)}}{z^{(0)}}-\beta\mu_j^{(1)}\biggr)^2\biggr]~.\label{eq:bose-k-c}
\end{align}
\end{subequations}

Accordingly, constraints~\eqref{eq:constr} yield the parameters $\xi^{(k)}$ and $z^{(k)}$ as a function of $\mu^{(k)}$ and $\{\mu_j^{(k)}\}_{j\ge 1}$. For $k\le 2$, we obtain the system
\begin{subequations}\label{eq:xi-k}
\begin{align}
N\xi^{(0)}&=\bigl[{z^{(0)}}^{-1}e^{\beta\mu^{(0)}}-1\bigr]^{-1}\nonumber\\
 &=N-\sum_{j=1}^\infty \bigl[{z^{(0)}}^{-1}e^{\beta\mu_j^{(0)}}-1\bigr]^{-1}~,\label{eq:xi-k-0}\\
N\xi^{(1)}&= \frac{{z^{(0)}}^{-1}e^{\beta\mu^{(0)}}}{\bigl[{z^{(0)}}^{-1}e^{\beta\mu^{(0)}}-1\bigr]^2}\biggl(\frac{z^{(1)}}{z^{(0)}}-\beta\mu^{(1)}\biggr)\nonumber\\
 &=-\sum_{j=1}^{\infty}\frac{{z^{(0)}}^{-1}e^{\beta\mu_j^{(0)}}}{\bigl[{z^{(0)}}^{-1}e^{\beta\mu_j^{(0)}}-1\bigr]^2}\biggl(\frac{z^{(1)}}{z^{(0)}}-\beta\mu_j^{(1)}\biggr)~,\label{eq:xi-k-1}
\end{align}
\begin{align} 
N\xi^{(2)}&=\frac{{z^{(0)}}^{-1}e^{\beta\mu^{(0)}}}{\bigl[{z^{(0)}}^{-1}e^{\beta\mu^{(0)}}-1\bigr]^2}\biggl[-\biggl(\frac{{z^{(1)}}^2}{{z^{(0)}}^2}-\frac{z^{(2)}}{z^{(0)}}-\beta\mu^{(1)}\frac{z^{(1)}}{z^{(0)}}+\beta\mu^{(2)}\nonumber\\
&\mbox{}\quad +{\textstyle \frac{1}{2}}\beta^2{\mu^{(1)}}^2\biggr)+\frac{{z^{(0)}}^{-1}e^{\beta\mu^{(0)}}}{{z^{(0)}}^{-1}e^{\beta\mu^{(0)}}-1}\biggl(\frac{z^{(1)}}{z^{(0)}}-\beta\mu^{(1)}\biggr)^2\biggr]~,\nonumber\\
&=-\sum_{j=1}^\infty \frac{{z^{(0)}}^{-1}e^{\beta\mu_j^{(0)}}}{\bigl[{z^{(0)}}^{-1}e^{\beta\mu_j^{(0)}}-1\bigr]^2}\biggl[-\biggl(\frac{{z^{(1)}}^2}{{z^{(0)}}^2}-\frac{z^{(2)}}{z^{(0)}}-\beta\mu_j^{(1)}\frac{z^{(1)}}{z^{(0)}}+\beta\mu_j^{(2)}\nonumber\\
&\mbox{}\quad +{\textstyle \frac{1}{2}}\beta^2{\mu_j^{(1)}}^2\biggr)+\frac{{z^{(0)}}^{-1}e^{\beta\mu_j^{(0)}}}{{z^{(0)}}^{-1}e^{\beta\mu_j^{(0)}}-1}\biggl(\frac{z^{(1)}}{z^{(0)}}-\beta\mu_j^{(1)}\biggr)^2\biggr]~.\label{eq:xi-k-2}
\end{align}
\end{subequations}
Thus, solving for $z^{(k)}$, we compute
\begin{subequations}\label{eq:xi-z-0}
\begin{align}
z^{(0)}&=e^{\beta\mu^{(0)}}\big[1+(N\xi^{(0)})^{-1}\big]^{-1}~,\label{eq:xi-z-0-a}\\
N\xi^{(0)}&=N-\sum_{j=1}^\infty \bigl\{[1+(N\xi^{(0)})^{-1}]e^{\beta(\mu_j^{(0)}-\mu^{(0)})}-1\bigr\}^{-1}~;\label{eq:xi-z-0-b}
\end{align}
\end{subequations}
\begin{subequations}\label{eq:xi-z-1}
\begin{align}
\frac{z^{(1)}}{z^{(0)}}&=\beta\mu^{(1)}+N\xi^{(1)}\,z^{(0)}e^{-\beta\mu^{(0)}}\big[{z^{(0)}}^{-1}e^{\beta\mu^{(0)}}-1\big]^2~,\label{eq:xi-z-1-a}\\
N\xi^{(1)}&=-\left\{1+z^{(0)}e^{-\beta\mu^{(0)}}\big[{z^{(0)}}^{-1}e^{\beta\mu^{(0)}}-1\big]^2\sum_{j=1}^{\infty}
\frac{{z^{(0)}}^{-1}e^{\beta\mu_j^{(0)}}}{\bigl[{z^{(0)}}^{-1}e^{\beta\mu_j^{(0)}}-1\bigr]^2}\right\}^{-1}\nonumber\\
&\mbox{}\quad \times
\sum_{j=1}^{\infty}\frac{{z^{(0)}}^{-1}e^{\beta\mu_j^{(0)}}}{\bigl[{z^{(0)}}^{-1}e^{\beta\mu_j^{(0)}}-1\bigr]^2}
\beta\big(\mu^{(1)}-\mu_j^{(1)}\big)~;\label{eq:xi-z-1-b}
\end{align}
\end{subequations}
\begin{align}
\frac{z^{(2)}}{z^{(0)}}&=\frac{{z^{(1)}}^2}{{z^{(0)}}^2}+z^{(0)}e^{-\beta\mu^{(0)}}\big({z^{(0)}}^{-1}e^{\beta\mu^{(0)}}-1\big)^2\big\{N\xi^{(2)}-N\xi^{(1)}\big[\beta\mu^{(1)}\nonumber\\
&\mbox{}\quad +N\xi^{(1)}\big({z^{(0)}}^{-1}e^{\beta\mu^{(0)}}-1\big)\big]\big\}+\beta\mu^{(2)}-{\textstyle\frac{1}{2}}\big(\beta\mu^{(1)}\big)^2~.~\label{eq:xi-z-2}
\end{align}
Equations~\eqref{eq:xi-z-0} should provide $(z^{(0)},N\xi^{(0)})$ as a function of $\mu^{(0)}$ and $\{\mu_j^{(0)}\}_{j\ge 1}$ via solving a nonlinear equation for $N\xi^{(0)}$. Equations~\eqref{eq:xi-z-1} then readily yield $(z^{(1)},N\xi^{(1)})$.
Equation~\eqref{eq:xi-z-2} can be used in the $\epsilon$-expansion of the second constraint in~\eqref{eq:constr} to obtain $(z^{(2)},N\xi^{(2)})$. Finally, $b_j^{(k)}$ are determined from appropriately expanding~\eqref{eq:bj-stat}, or, alternatively, by requiring that $\langle \Phi,\phi_j\rangle=0$ be satisfied order by order (see Appendix).

\subsection{Effective equations of motion up to order $k=2$}
\label{subsec:effective}

We proceed to homogenize PDEs~\eqref{eq:casc-Phi} and~\eqref{eq:casc-phj} in  view of~\eqref{eq:per-micro} and~\eqref{eq:s-dens-k}--\eqref{eq:xi-z-2} under the assumption that $A(y)$ has zero mean, $\langle A\rangle=0$. The resulting equations for the slowly-varying parts of $\Phi^{(k)}$ and $\phi_j^{(k)}$ with $k=0,\,1,\,2$ are displayed in~\eqref{eq:hom-Ph-ph0}, \eqref{eq:hom-Ph-ph1} and~\eqref{eq:hom-Ph-ph2}.
 
By~\eqref{eq:casc-Phi0}, \eqref{eq:casc-phj0} and Lemma~\ref{lem:Fred}, we trivially have $\langle S_{\Phi}^{(0)}\rangle=0=\langle S_{\phi,j}^{(0)}\rangle$ and conclude that
\begin{equation*}
\Phi^{(0)}(\bx, x)=f^0(x)~,\quad \phi_j^{(0)}(\bx ,x)=f_j^0(x)~,
\end{equation*}
by which the zeroth-order densities are $\varrho_{\ell}^{(0)}(\bx,x)=\rho_{\ell}^0(x)$, for $\ell=s,\,n$. Similarly, the trivially satisfied equations $\langle S_{\Phi}^{(1)}\rangle=0=\langle S_{\phi,j}^{(1)}\rangle$ yield
\begin{equation}
\Phi^{(1)}(\bx, x)=f^1(x)~,\quad \phi_j^{(1)}(\bx ,x)=f_j^1(x)~;
\end{equation}
thus, $\varrho_{\ell}^{(1)}(\bx, x)=\rho_{\ell}^1(x)$. In regard to the density coefficients $\varrho_\ell^{(k)}$ ($k=0,\,1$), recall~\eqref{eq:s-dens-k-a}, \eqref{eq:s-dens-k-b}, \eqref{eq:n-dens-k-a} and~\eqref{eq:n-dens-k-b}.

By~\eqref{eq:casc-Phi2} and~\eqref{eq:casc-phj2}, setting $\langle S_\Phi^{(2)}\rangle=0=\langle S_{\phi,j}^{(2)}\rangle$ entails the homogenized equations of motion for $f^0(x)$ and $f_j^0(x)$, with $\Vert f^0\Vert^2=N$ and $\Vert f_j^0\Vert=1$:
\begin{subequations}\label{eq:hom-Ph-ph0}
\begin{align}
& \{-\Delta_x+V_e(x)+g_0[\varrho_s^0(x)+2\varrho_n^0(x)]\} f^0(x)=\mu^{(0)}f^0(x)~,\label{eq:hom-Ph0}\\
& \{-\Delta_x+V_e(x)+2g_0[\varrho_s^0(x)+\varrho_n^0(x)]\} f_j^0(x)-b_j^{(0)} f^0(x)=\mu_j^{(0)}f_j^0(x)~,\label{eq:hom-ph0}
\end{align}
\end{subequations}
where $b_j^{(0)}=N^{-1}\xi^{(0)}g_0\langle {(f^0)}^3, f_j^0\rangle$. The eigenvalues $\mu^{(0)}$ and $\mu_j^{(0)}$ of course form part of the solution. By Lemma~\ref{lem:Fred}, the admissible second-order coefficients of the solution read
\begin{subequations}\label{eq:Phi-ph2}
\begin{align}
\Phi^{(2)}(\bx,x)&=-g_0[\rho_s^0(x)+2\rho_n^0(x)]f^0(x)[(-\Delta_{\bx})^{-1}A(\bx)]+f^2(x)~,\label{eq:Phi-2}\\
\phi_j^{(2)}(\bx,x)&=-2g_0[\rho_s^0(x)+\rho_n^0(x)]f^0_j(x)[(-\Delta_{\bx})^{-1}A(\bx)]+f^2_j(x)~,\label{eq:phij-2}
\end{align}
\end{subequations}
where the (slowly-varying) functions $f^2(x)$ and $f_j^2(x)$ should be determined. 
By~\eqref{eq:s-dens-k-c} and~\eqref{eq:n-dens-k-c}, for $k=2$, the second-order coefficients for the densities read
\begin{subequations}\label{eq:hom-dens2}
\begin{align}
\varrho_s^{(2)}(\bx,x)&=\rho_s^2(x)-g_0\bar\rho_s^2(x) [(-\Delta_{\bx})^{-1}A(\bx)]~,\label{eq:hom-dens2-s}\\
\varrho_n^{(2)}(\bx,x)&=\rho_n^2(x)-g_0\bar\rho_n^2(x) [(-\Delta_{\bx})^{-1}A(\bx)]~,\label{eq:hom-dens2-n}
\end{align}
where
\begin{align*}
\rho_s^2(x)&=\xi^{(2)}f^0(x)^2+2\xi^{(1)}f^1(x)f^0(x)+\xi^{(0)}[2f^0(x)f^2(x)+f^1(x)^2]~,\\
\bar\rho_s^2(x)&=2\rho_s^0(x)[\rho_s^0(x)+2\rho_n^0(x)]~,\\
\rho_n^2(x)&=\sum_{j=1}^\infty \bigl\{n_j^{(2)}f_j^0(x)^2+2n_j^{(1)}f_j^1(x)f_j^0(x)+n_j^{(0)}[2f_j^2(x)f_j^0(x)+f_j^1(x)^2]\bigr\}~,\\
\bar\rho_n^2(x)&=4[\rho_s^0(x)+\rho_n^0(x)]\rho_n^0(x)~.
\end{align*}
\end{subequations}

To find equations of motion for $f^k$ and $f_j^k$ with $k=1,\,2$, we need to apply Lemma~\ref{lem:Fred} to the next two higher orders. By~\eqref{eq:casc-Phi3} and~\eqref{eq:casc-phj3}, the requirement $\langle S_\Phi^{(3)}\rangle=0=\langle S_{\phi,j}^{(3)}\rangle$ leads to
\begin{subequations}\label{eq:hom-Ph-ph1}
\begin{align}
&\{-\Delta_x+V_e(x)+g_0[\rho_s^0(x)+2\rho_n^0(x)]-\mu^{(0)}\}f^1(x)\nonumber\\
&\mbox{}\qquad =\{\mu^{(1)}-g_0[\rho_s^1(x)+2\rho_n^1(x)]\}f^0(x)~,\label{eq:hom-Ph1}\\
&\{-\Delta_x+V_e(x)+2g_0[\rho_s^0(x)+\rho_n^0(x)]-\mu_j^{(0)}\}f^1_j(x)-b_j^{(0)}f^1-b_j^{(1)}f^0\nonumber\\
&\mbox{}\qquad =\{\mu_j^{(1)}-2g_0[\rho_s^1(x)+\rho_n^1(x)]\}f_j^0(x)~,\label{eq:hom-ph1}
\end{align}
\end{subequations}
where $\mu^{(1)}$ and $\mu_j^{(1)}$ are subject to $\langle f^1, f^0\rangle =0$ and $\langle f_j^1, f_j^0\rangle=0$; and $b_j^{(1)}$ is found from imposing $\langle f^1, f_j^0\rangle +\langle f^0, f_j^1\rangle =0$ (see Appendix for a derivation):
\begin{align*}
b_j^{(1)}&=\big\{\langle f^1,\rho_s^0f_j^0\rangle+\langle f_j^1,\rho_s^0f^0\rangle+\langle f^0,\rho_s^1 f_j^0\rangle\big\}\nonumber\\
&=N^{-1}g_0\big\{ \xi^{(0)}\langle (f^0)^3,f_j^1\rangle+3\xi^{(0)}\langle f^1,(f^0)^2f_j^0\rangle+\xi^{(1)}\langle (f^0)^3,f_j^0\rangle\big\}~.
\end{align*}
By Lemma~\ref{lem:Fred}, the third-order coefficients in the $\epsilon$-expansion of $\Phi$ and $\phi_j$ read
\begin{subequations}\label{eq:Phi-ph3}
\begin{align}
\Phi^{(3)}(\bx,x)&=
-2g_0 \nabla_x\big[\big(\rho_s^0(x)+2\rho_n^0(x)\big)f^0(x)\big]\cdot\nabla_{\bx} (\Delta_{\bx}^{-2}A)\nonumber\\
&\mbox{}\quad -g_0\big[\big(\rho_s^0(x)+2\rho_n^0(x)\big)f^1(x)+\big(\rho_s^1(x)+2\rho_n^1(x)\big)f^0(x)\big]\nonumber\\
&\mbox{}\qquad \times\{(-\Delta_{\bx})^{-1}A\}+f^3(x)~,\label{eq:Phi-3}\\
\phi_j^{(3)}(\bx,x)&=-4 g_0 \nabla_x\big[\big(\rho_s^0(x)+\rho_n^0(x)\big)f_j^0(x)\big]\cdot\nabla_{\bx} (\Delta_{\bx}^{-2}A)\nonumber\\
&\mbox{}\quad -2g_0\big[\big(\rho_s^0(x)+\rho_n^0(x)\big)f_j^1(x)+\big(\rho_s^1(x)+\rho_n^1(x)\big)f_j^0(x)\big]\nonumber\\
&\mbox{}\qquad \times\{(-\Delta_{\bx})^{-1}A\}+f_j^3(x)~.\label{eq:phj-3}
\end{align}
\end{subequations}

We continue our homogenization program, proceeding to the next higher order. By~\eqref{eq:casc-Phi4} and~\eqref{eq:casc-phj4}, setting $\langle S_{\Phi}^{(4)}\rangle=0=\langle S_{\phi,j}^{(4)}\rangle$ yields the following PDEs.
\begin{subequations}\label{eq:hom-Ph-ph2}
\begin{align}
&\{-\Delta_x+V_e(x)+g_0[\rho_s^0(x)+2\rho_n^0(x)]-\mu^{(0)}\}f^2(x)\nonumber\\
&=\big\{-g_0\big(\rho_s^1(x)+2\rho_n^1(x)\big)+\mu^{(1)}\big\}f^1(x)\nonumber\\
&\mbox{}\quad +\big\{g_0^2\Vert A\Vert^2_{-1} \big(\rho_s^{0}(x)+2\rho_n^0(x)\big)^2-g_0^2\Vert A\Vert^{2}_{-1}\big(\bar\rho_s^2(x)+2\bar\rho_n^2(x)\big)\nonumber\\
&\mbox{}\quad -g_0\big(\rho_s^2(x)+2\rho_n^2(x)\big)+\mu^{(2)}\big\} f^0(x)~,\label{eq:hom-Ph2}
\end{align}
\begin{align}
&\{-\Delta_x+V_e(x)+2g_0[\rho_s^0(x)+\rho_n^0(x)]-\mu_j^{(0)}\}f_j^2(x)-b_j^{(0)}f^2-b_j^{(1)}f^1-b_j^{(2)}f^0\nonumber\\
&=\big\{-2g_0\big(\rho_s^1(x)+\rho_n^1(x)\big)+\mu_j^{(1)}\big\}f_j^1(x)\nonumber\\
&\mbox{}\quad +\big\{4g_0^2\Vert A\Vert^2_{-1} \big(\rho_s^{0}(x)+\rho_n^0(x)\big)^2-2g_0^2\Vert A\Vert^{2}_{-1}\big(\bar\rho_s^2(x)+\bar\rho_n^2(x)\big)\nonumber\\
&\mbox{}\quad -2g_0\big(\rho_s^2(x)+\rho_n^2(x)\big)+\mu_j^{(2)}\big\} f_j^0(x)~,\label{eq:hom-ph2}
\end{align}
\end{subequations}
where $\Vert A\Vert_{-1}$ is the $H^{-1}$-norm of the $1$-periodic, fast-varying function $A(\bx)$; recall that $\langle A\rangle =0$. The factor
$b_j^{(2)}$ is determined directly from~\eqref{eq:bj-stat} which is consistent with imposing the condition $\langle f^0, f_j^2\rangle+\langle f^1, f_j^1\rangle+\langle f^2,f_j^0\rangle=0$; see~\eqref{eq:bj-2-stat} in Appendix. The parameters $\mu^{(2)}$ and $\mu_j^{(2)}$ are subject to $\langle f^0,\Phi^{(2)}\rangle +\langle f^1,f^1\rangle +\langle \Phi^{(2)},f^0\rangle=0$ and $\langle f_j^0,\phi_j^{(2)}\rangle +\langle f^1_j,f^1_j\rangle +\langle \phi^{(2)}_j,f^0_j\rangle=0$ which, by Lemma~\ref{lem:A1} of Appendix, yield $\Vert f^1\Vert^2+2\langle f^0,f^2\rangle=0$ and  $\Vert f^1_j\Vert^2+2\langle f^0_j,f^2_j\rangle=0$, respectively.
\medskip

The main results  of our formal procedure are summarized as follows. 
\medskip

\begin{proposition}
Consider governing equations~\eqref{eq:phis-stat} for $\Phi(x)$ and $\phi_j(x)$ ($j\in\mathbb{N}_*$) under spatially-varying scattering length~\eqref{eq:per-micro}.  Two-scale expansion~\eqref{eq:2-scale} entails $\Phi^{(0)}(\bx, x)=f^0(x)$ and $\Phi^{(1)}(\bx,x)=f^1(x)$, which depend only on the slow variable, while
$\Phi^{(2)}(\bx,x)$ and $\phi_j^{(2)}(\bx,x)$ are given by~\eqref{eq:Phi-ph2}. The slow-varying functions $f^k(x)$ and $f_j^k(x)$ ($k=0,\,1,\,2$) satisfy~\eqref{eq:hom-Ph-ph0}, \eqref{eq:hom-Ph-ph1} and~\eqref{eq:hom-Ph-ph2}. In these equations, the parameters $\xi^{(k)}$ and $z^{(k)}$ appearing in the density coefficients $\rho_s^{(k)}$ and $\rho_n^{(k)}$ are functions of $\mu^{(k)}$ and $\mu_j^{(k)}$ according to~\eqref{eq:xi-z-0}--\eqref{eq:xi-z-2} with~\eqref{eq:constr}.
\end{proposition}


\subsection{Energy of Boson system}
\label{subsec:sys-en}

We conclude the homogenization program by discussing the total energy~\eqref{eq:total-en}. 
The requisite energy parameters are $E$ and $E_j$; cf.~section~\ref{sec:stat}. By~\eqref{eq:mus-def}, we deduce that
\begin{equation}
E=E^{(0)}+\epsilon E^{(1)}+\epsilon^2 E^{(2)}+\ldots\,,\quad
E_j=E_j^{(0)}+\epsilon E_j^{(1)}+\epsilon^2 E_j^{(2)}+\ldots\,,
\end{equation}
where 
\begin{align}
E^{(0)}&=\mu^{(0)}-{\textstyle\frac{1}{2}}{\xi^{(0)}}^2\zeta^{(0)}~,\nonumber\\
E^{(1)}&=\mu^{(1)}-{\textstyle\frac{1}{2}}\big[{\xi^{(0)}}^2\zeta^{(1)}+2\xi^{(0)}\xi^{(1)}\zeta^{(0)}\big]~,\nonumber\\
E^{(2)}&=\mu^{(2)}-{\textstyle\frac{1}{2}}\big[{\xi^{(0)}}^2\zeta^{(2)}+2\xi^{(0)}\xi^{(1)}\zeta^{(1)}+\big(2\xi^{(0)}\xi^{(2)}+{\xi^{(1)}}^2\big)\zeta^{(0)}\big]~;
\end{align}
and similarly for $E_j$ via replacement of $\mu^{(k)}$ by $\mu_j^{(k)}$. In the above, $\zeta^{(k)}$ result from the $\epsilon$-expansion of the static version of~\eqref{eq:zeta-def} via Lemma~\ref{lem:A2} of Appendix; viz.,
\begin{align*}
\zeta&=\zeta^{(0)}+\epsilon\zeta^{(1)}+\epsilon^2 \zeta^{(2)}+\ldots\,\nonumber\\
&=N^{-1}g_0\big\{ \Vert (f^0)^2\Vert^2+4\epsilon \langle f^1,(f^0)^3\rangle+\epsilon^2\big[4\langle f^2,(f^0)^3\rangle +6\Vert f^1f^0\Vert^2\nonumber\\
&\mbox{}\qquad -4g_0\Vert A\Vert_{-1}^2\, \Vert \sqrt{\rho_s^0+2\rho_n^0}(f^0)^2\Vert^2\big]\big\}~.  
\end{align*}

Consequently, by~\eqref{eq:total-en} the total energy reads $\mathcal E=\mathcal E^{(0)}+\epsilon \mathcal E^{(1)}+\epsilon^2 \mathcal E^{(2)}+\ldots\,$, where
\begin{align}\label{eq:tot-en-coeffs}
\mathcal E^{(0)}&= N\xi^{(0)}+\sum_j n_j^{(0)}E_j^{(0)}-2g_0 \langle\rho_s^0,\rho_n^0\rangle-g_0\Vert\rho_n^0\Vert^2~,\nonumber\\
\mathcal E^{(1)}&=N\big( \xi^{(0)}E^{(1)}+\xi^{(1)}E^{(0)}\big)+\sum_j \big( n_j^{(0)}E_j^{(1)}+n_j^{(1)}E_j^{(0)}\big)\nonumber\\
&\mbox{}\quad -2g_0\big( \langle \rho_n^1,\rho_s^0\rangle+\langle \rho_s^1,\rho_n^0\rangle+\langle \rho_n^0,\rho_n^1\rangle\big)~,\nonumber\\
\mathcal E^{(2)}&= N\big(\xi^{(0)}E^{(2)}+\xi^{(1)}E^{(1)}+\xi^{(2)}E^{(0)}\big)\nonumber\\
&\mbox{}\quad 
+\sum_j \big(n_j^{(0)}E_j^{(2)}+n_j^{(1)}E_j^{(1)}+n_j^{(2)}E_j^{(0)}\big)\nonumber\\
&\mbox{}\quad -2g_0\big\{ \langle\rho_n^0,\rho_s^2\rangle+\langle\rho_s^1,\rho_n^1\rangle +\langle\rho_s^0,\rho_n^2\rangle+\langle \rho_n^2,\rho_n^0\rangle +{\textstyle\frac{1}{2}}
\Vert\rho_n^1\Vert^2\nonumber\\
&\mbox{}\quad -g_0\Vert A\Vert^2 \big(\langle \rho_n^0,\bar\rho_s^2\rangle+\langle\rho_s^0,\bar\rho_n^2\rangle+\langle \rho_n^0,\bar\rho_n^2\rangle\big)\big\}~.
\end{align}

\section{Discussion and conclusion}
\label{sec:discussion}

In this article, we introduced a formalism for Bose-Einstein condensates in an inhomogeneous scattering environment at sufficiently low, finite temperatures. By starting with the Hamiltonian description of $N$ repulsively interacting Bosons forming a dilute gas in a trap, we formally derived mean field evolution equations for the condensate wave function and the single-particle wave functions of thermally excited states. Our main assumption is that the temperature is finite but lies sufficiently below the phase transition point. In our model, the scattering length, which expresses the strength of the pairwise particle interactions, is phenomenologically included in the Hamiltonian as a spatially varying, periodic function of subscale $\epsilon$.  

First, we reduced the many-body problem to lower-dimensional evolution PDEs on the basis of a perturbative treatment of the many-body Hamiltonian in Fock space. This methodology is distinctly different from variational approaches for stationary settings, e.g.~\cite{goldmanetal,huseetal}; and enabled us to derive mean field laws for {\em time-dependent} one-particle wave functions. Our analysis is a direct extension to finite temperatures of techniques applied to the case with zero temperature~\cite{Wu98,Grillakis10}. In the present work, we restricted attention to formal derivations, by invoking uncontrolled ansatz~\eqref{eq:Psi-tp} for the $N$-body Schr\"odinger state vector. The particle occupation numbers, $n_j$, for thermally excited states are treated as parameters in the evolution laws.

Second, we homogenized the resulting PDEs up to second order in $\epsilon$ for stationary states. This procedure relies on a scale separation for the condensate wave function and the one-particle wave functions of the thermally excited states. As a result, we obtained a system of $\epsilon$-independent equations of motion for the slowly-varying parts of the single-particle wave functions.  In these effective equations, the oscillatory zero-mean ingredient, $A$, of the scattering length contributes terms proportional to its $H^{-1}$ norm squared, analogously to the case with zero temperature~\cite{Margetis12}. 

There is a number of issues that remain unresolved within our treatment. In  optical traps, the rapid spatial variation of the scattering length is driven by the external potential, $V_e(x)$, set by appropriate laser fields~\cite{chin10}. This suggests that, in realistic settings, this $V_e(x)$ should also vary spatially with the same subscale, $\epsilon$. The joint effect of a rapidly varying $V_e$ and the periodic scattering length, $a$, is the subject of work in progress. The current treatment, placing emphasis on the extraction of low-dimensional laws of motion, has not addressed the issue of approximate solutions to these equations, e.g., in the Thomas-Fermi regime~\cite{chouyang}; in this regime, it is possible to derive an approximate description for the superfluid density combined with an approximation for the normal fluid density. This task is left for future work. The spatially dependent scattering length is introduced as an ad hoc parameter in the microscopic Hamiltonian.	The emergence of such a parameter from the limiting procedure lies beyond our present scope.	The finite-temperature approach here introduces the occupation numbers, $n_j$, of one-particle thermally excited states as given parameters in the quantum dynamics; the derivation of these parameters was not addressed. The present formalism leaves out two effects: quantum depletion of the condensate due to pair excitation, which calls for modifying ansatz~\eqref{eq:Psi-tp}; and the effect of dissipation due to couplings of the condensate with the environment. The effect of quantum depletion, brought about by the operator $\mathcal V_{21}$ of~\eqref{eq:V21}, should be tractable through the introduction of the suitable pair-excitation kernel into the ansatz for the many-body Schr\"odinger state vector~\cite{Wu98}. This approach, which transcends the usual mean field limit, is a promising direction of research. Finally, we believe that the approximations for the $N$-body Hamiltonian are valid over some moderate time scale; the precise characterization of this time regime has been left unresolved.

\appendix

\section{On the computation of $b_j^{(k)}$, for $k=1,\,2$}
In this appendix, we compute the coefficients $b_j^{(k)}$ in the $\epsilon$-expansion of formula~\eqref{eq:bj-stat} with $k=1,\,2$.
For this purpose, we need a few results pertaining to classical asymptotics of integrals of highly oscillatory functions~\cite{Margetis12}.

\begin{lemma}\label{lem:A1} 
Consider the integrable function $\varphi : \mathbb{R}^d \rightarrow \mathbb{R}$ and the bounded,
1-periodic $P : \mathbb{T}^d \rightarrow \mathbb{R}$ with $\langle P\rangle = 0$ ($d\ge 1$). 
Suppose that $\varphi$ has
$m$ summable derivatives, $m\in \mathbb{N}_*$, which vanish at infinity; then,
\begin{equation}\label{eq:A1}
 \int_{\mathbb{R}^d} P\biggl(\frac{x}{\epsilon}\biggr)\, \varphi(x)\,\dx=\mathcal O(\epsilon^m)\quad\mbox{as}\ \epsilon\downarrow 0~.
\end{equation}
In the special case with smooth $\varphi$ ($\varphi\in C^\infty(\mathbb{R}^d)$), 
\begin{equation}
\lim_{\epsilon\downarrow 0}\left\{\epsilon^{-k}\int_{\mathbb{R}^d} P\biggl(\frac{x}{\epsilon}\biggr)\, \varphi(x)\,\dx\right\}=0~,
\end{equation}
for any $k\in\mathbb{N}$.
\end{lemma}

The proof of this lemma can be established via successive integrations by parts~\cite{Margetis12}, and is omitted here. 

More generally, for 1-periodic and bounded  $P : \mathbb{R}^d \rightarrow \mathbb{R}$ with $\langle P\rangle\neq 0$, one can write 
$P(y)=\langle P\rangle +\bar P(y)$, where $\bar P(y)$ satisfies the hypotheses of Lemma~\ref{lem:A1}; particularly, $\langle P\rangle =0$. Thus, we reach the following conclusion.

\begin{lemma}\label{lem:A2}
Consider the integrable $\varphi : \mathbb{R}^d \rightarrow \mathbb{R}$ and the bounded,
1-periodic $P : \mathbb{T}^d \rightarrow \mathbb{R}$ with finite $\langle P\rangle$, $\langle P\rangle \neq 0$ ($d\ge 1$). 
Suppose that $\varphi$ has
$m$ summable derivatives, $m \in \mathbb{N}_*$, which vanish at infinity; then,
\begin{equation}
\int_{\mathbb{R}^d}P\biggl(\frac{x}{\epsilon}\biggr)\, \varphi(x)\,\dx=\langle P\rangle\int_{\mathbb{R}^d}\varphi(x)\,\dx+\mathcal O(\epsilon^m)\quad \mbox{as}\ \epsilon\downarrow 0~.
\end{equation}
\end{lemma}

We proceed to describe the computation of $b_j^{(k)}$ ($k=1,\,2$).
By inspection of~\eqref{eq:bj-stat} along with the formulas for $\Phi^{(k)}$ and $\phi_j^{(k)}$ provided in section~\ref{subsec:effective}, we need to use Lemma~\ref{lem:A1} for the bounded (and smooth) $P(y)=A(y)$ and Lemma~\ref{lem:A2} for the bounded $P(y)=A(y)[(-\Delta_y)^{-1}A(y)]$, $y=x/\epsilon$. On the other hand,  $\varphi(x)$ contains products of $f^k$'s and $f_j^k$'s, each of which is assumed to be smooth. Note that 
\begin{equation*}
\langle A[(-\Delta)^{-1}A]\rangle=\langle (-\Delta)^{-1}A, A\rangle=\Vert A\Vert_{-1}^2~,
\end{equation*}
which is the $H^{-1}$-norm squared of the (1-periodic, zero-mean) $A(y)$.

Without further ado, by $\Phi\sim f^0+\epsilon f^1+\epsilon^2 \Phi^{(2)}$ and $\phi_j\sim f_j^0+\epsilon f_j^1+\epsilon^2 \phi_j^{(2)}$, \eqref{eq:bj-stat} yields
\begin{align}\label{eq:bj-1-stat}
b_j^{(1)}&= N^{-1}g_0 \lim_{\epsilon\downarrow 0}\int_{\mathbb{R}^3}[1+A(x/\epsilon)]\,\big\{\xi^{(0)}\big[f_j^1(x)f^0(x)^3+3f^0(x)^2f^1(x)f_j^0(x)\big]\nonumber\\
&\mbox{}\quad +\xi^1 f^0(x)^3f_j^0(x)\big\}\nonumber\\
&= N^{-1}g_0 \big\{\xi^{(0)}\big[\langle (f^0)^3,f_j^1\rangle+3\langle (f^0)^2f_j^0, f^1\rangle\big]+\xi^{(1)}\langle (f^0)^3, f_j^0\rangle\big\}
\end{align}
and
\begin{align}\label{eq:bj-2-stat}
b_j^{(2)}&= {\textstyle\frac{1}{2}}N^{-1}g_0 \lim_{\epsilon\downarrow 0}\frac{\partial^2}{\partial \epsilon^2}\biggl\{(\xi^{(0)}+\epsilon\xi^{(1)}+\epsilon^2\xi^{(2)})\int_{\mathbb{R}^3}[1+A(x/\epsilon)]\nonumber\\
&\mbox{}\quad \times \big[ f^0(x)+\epsilon f^1(x)+\epsilon^2 \Phi^{(2)}(x/\epsilon,x)\big]^3\nonumber\\
&\mbox{}\quad \times \big[f_j^0(x)+\epsilon f_j^1(x)+\epsilon^2 \phi_j^{(2)}(x/\epsilon,x)\big]\biggr\}\nonumber\\
&=N^{-1}g_0\big\{\xi^{(0)}\big[\langle (f^0)^3,f_j^2\rangle+3\big(\langle (f^0)^2,f^1f_j^1\rangle+\langle (f^1)^2,f^0f_j^0\rangle\nonumber\\
&\mbox{}\quad +\langle (f^0)^2f_j^0,f^2\rangle\big)-g_0\Vert A\Vert^2\langle f_j^0,(5\rho_s^0+8\rho_n^0)(f^0)^3\rangle\big]\nonumber\\
&\mbox{}\quad +\xi^{(1)}\big[ \langle (f^0)^3,f_j^1\rangle +3 \langle (f^0)^2f_j^0,f^1\rangle\big]+\xi^{(2)}\langle (f^0)^3,f_j^0\rangle\big\}~.
\end{align}

{\bf Acknowledgments.}
The author is grateful to Professor T.~T. Wu for his valuable advice without which this article could not have been written. He also wishes to thank Professor W. Bao for bringing Ref.~\cite{Baoetal13} to
the author's attention, and Professor M.~G. Grillakis for useful remarks on the manuscript. 


\begin{thebibliography}{[50]}


\bibitem{Cornell95}{\sc M.~H. Anderson, J.~R. Ensher, M.~R. Matthews, C.~E.
Wieman, and E.~A. Cornell}, {\em Observation of Bose-Einstein condensation in a dilute atomic vapor}, Science, 269 (1995), pp.\ 198--201.

\bibitem{Baoetal13}{\sc W. Bao and Y. Cai}, {\em Mathematical theory and numerical methods for Bose-Einstein condensation},
Kinet.\ Relat.\ Mod., 6 (2013), pp.\ 1--135.

\bibitem{Baoetal04}{\sc W. Bao, L. Pareschi, and P.~A. Markowich}, {\em Quantum kinetic theory: modelling and numerics for Bose-Einstein condensation}, in Modeling and Computational Methods for Kinetic Equations, in: Modeling and Simulation in Science, Engineering and Technology,
P. Degond, L. Pareschi, NS G. Russo, Eds., Birkh\"auser, Boston, 2004, pp.\ 287--320.


\bibitem{hom-book}
{\sc A. Bensoussan, J.~L. Lions, and G.~C. Papanicolaou},
{\em Asymptotic Analysis of Periodic Structures}, North Holland, Amsterdam, 1978.

\bibitem{Berezin}{\sc F.~A. Berezin}, {\em The Method of Second Quantization},
Academic Press, New York, NY, 1966.

\bibitem{BlakieDavis05}{\sc P. Blair Blakie and M.~J. Davis}, {\em Projected Gross-Pitaevskii equation for harmonically
confined Bose gases at finite temperatures}, Phys.\ Rev., A (2005), 063608.


\bibitem{Carretero08}{\sc R. Carretero-Gonz\'alez, D.~J. Frantzeskakis, and P.~G. Kevrekidis},
{\em Nonlinear waves in Bose-Einstein condensates:
Physical relevance and mathematical techniques}, Nonlinearity, 21 (2008), pp.\ R139--R202.

\bibitem{Carusottoetal13}{\sc I. Carusotto and C. Ciuti}, {\em Quantum fluids of light}, Rev.\ Mod.\ Phys., 85 (2013), pp.\ 299--366.

\bibitem{chin10}{\sc C. Chin, R. Grimm, P. Julienne, and E. Tiesinga},
{\em Feshbach resonances in ultracold gases}, Rev.\ Mod.\ Phys., 82 (2010), pp.\ 1225--1286.

\bibitem{chouyang}{\sc T.~T. Chou, C.~N. Yang, and L.~H. Yu}, {\em Bose-Einstein condensation of atoms in a trap}, Phys.\ Rev.\ A, 53
(1996), pp.\ 4257--4259.

\bibitem{Cornelletal02}{\sc E.~A. Cornell and C.~E. Wieman},
{\em Nobel lecture: Bose-Einstein condensation in a dilute gas, the first 70 years and some recent experiments},
Rev.\ Mod.\ Phys., 74 (2002), pp.\ 875--893.

\bibitem{Cornish00}{\sc S.~L. Cornish, N.~R. Claussen, J.~L. Roberts, E.~A. Cornell, and
C.~E. Wieman}, {\em Stable ${}^{85}${\rm Rb} Bose-Einstein condensates with widely tunable interactions},
Phys.\ Rev.\ Lett., 85 (2000), pp.\ 1795--1798.


\bibitem{Croninetal09}{\sc A.~D. Cronin, J. Schmiedmayer, and D.~E. Pritchard}, {\em Optics and interferometry with atoms and molecules},
Rev.\ Mod.\ Phys.,  81 (2009), pp.\ 1051--1129.

\bibitem{Ketterle95}{\sc K.~B. Davis, M.-O. Mewes, M.~R. Andrews, N.~J. van
Druten, D.~S. Durfee, D.~M. Kurn, and W. Ketterle}, {\em Bose-Einstein condensation in a gas of sodium atoms}, Phys.\ Rev.\ Lett., 75 (1995), pp.\ 3969--3973.

\bibitem{Davisetal01}{\sc M.~J. Davis, S.~A. Morgan, and K. Burnett},
{\em Simulations of Bose fields at finite temperature}, Phys.\ Rev.\ Lett., 87 (2001),  160402.

\bibitem{dorreetal}{\sc P. D\"orre, H. Haug and D.~B. Tran Thoai}, {\em Condensate theory versus pairing theory for degenerate Bose systems},
J.\ Low Temp.\ Phys., 35 (1979), pp.\ 465--485.

\bibitem{Elgart06}{\sc A. Elgart, L. Erd\H{o}s, B. Schlein, and H.-T. Yau}, {\em Gross-Pitaevskii equation
as the mean field limit of weakly coupled Bosons}, Arch.\ Rat.\ Mech.\ Anal., 179 (2006), pp.\ 265--283.

\bibitem{Erdos06}
{\sc L. Erd\H{o}s, B. Schlein, and H.-T. Yau}, {\em Derivation of the Gross-Pitaevskii
hierarchy for the dynamics of Bose-Einstein condensate}, Comm.\ Pure Appl.\ Math., 59 (2006), pp.\ 1659--1741.

\bibitem{Erdos07}{\sc L. Erd\H{o}s, B. Schlein, and H.-T. Yau}, {\em Rigorous derivation of the Gross-Pitaevskii equation},
Phys.\ Rev.\ Lett., 98 (2007), 040404.

\bibitem{Fibich06}{\sc G. Fibich, Y. Sivan, and M.~I. Weinstein}, {\em Bound states of nonlinear Schr\"odinger equations with a periodic nonlinear microstructure},
Physica D, 217 (2006), pp.\ 31--57.

\bibitem{Giorgini97}{\sc S. Giorgini, L.~P. Pitaevskii, and S. Stringari}, {\em Thermodynamics of a trapped Bose gas}, J.\ Low Temp.\ Phys., 109
(1997), pp.\ 309--355.


\bibitem{goldmanetal}{\sc V.~V. Goldman, I.~F. Silvera, and A.~J. Leggett}, {\em Atomic hydrogen in an inhomogeneous magnetic field: Density profile and Bose-Einstein
condensation}, Phys.\ Rev.\ B, 24 (1981), pp.\ 2870--2873.

\bibitem{Grillakis10}{\sc M.~G. Grillakis, M. Machedon, and D. Margetis},
{\em Second-order corrections to mean field evolution of weakly interacting Bosons. I.}, Commun.\ Math.\ Phys., 294 (2010), pp.\ 273--301.

\bibitem{Gross61}{\sc E.~P. Gross}, {\em Structure of a quantized vortex in boson systems},
Nuovo Cim., 20 (1961), pp.\ 454--477.

\bibitem{Hagley99}{\sc E.~W. Hagley, L. Deng, M. Kozuma, J. Wen, K. Helmerson, S.~L. Rolston, and W.~D. Phillips},
{\em A well-collimated quasi-continuous atom laser}, Science, 283 (1999), pp.\ 1706--1709.

\bibitem{Huang-book}K. Huang, {\em Statistical Mechanics}, John Wiley, New York, NY, 1987.

\bibitem{HuangYang57}{\sc K. Huang and C.~N. Yang}, {\em Quantum-mechanical many-body problem with hard-sphere interaction},
Phys.\ Rev., 105 (1957), pp.\ 767--775.

\bibitem{huseetal}{\sc D.~A. Huse and E.~D. Siggia}, {\em The density distribution of a weakly interacting Bose gas in an external potential},
J.\ Low Temp.\ Phys., 46 (1982), pp.\ 137--149.


\bibitem{Ketterle02}
{\sc W. Ketterle}, {\em Nobel lecture: When atoms behave as waves: Bose-Einstein condensation and the atom laser},
Rev.\ Mod.\ Phys., 74 (2002), pp.\ 1131--1151.

\bibitem{leeetal}{\sc T.~D. Lee, K. Huang, and C.~N. Yang}, {\em Eigenvalues and eigenfunctions of a Bose system
of hard spheres}, Phys.\ Rev., 106 (1957), pp.\ 1135--1145.

\bibitem{leeyang58}{\sc T.~D. Lee and C.~N. Yang}, {\em Low-temperature behavior of a dilute Bose system of hard spheres. I. Equilibrium properties},
Phys.\ Rev., 112 (1958), pp.\ 1419--1429.

\bibitem{leeyang59}{\sc T.~D. Lee and C.~N. Yang}, {\em Low-temperature behavior of a dilute Bose system of hard spheres. II. Nonequilibrium properties},
Phys.\ Rev., 113 (1959), pp.\ 1406--1413.

\bibitem{London}
{\sc F. London}, {\em Superfluids, Vol. II: Macroscopic Theory of Superfluid Helium}, Dover, New York, NY, 1964.

\bibitem{Margetis12}{\sc D. Margetis}, {\em Bose-Einstein condensation beyond mean field: Many-body bound state of periodic microstructure},
Multiscale Model.\ Simul., 10 (2012), pp.\ 383--417.

\bibitem{MargetisThesis}{\sc D. Margetis}, {\em Studies in Classical Electromagnetic Radiation and Bose-Einstein Condensation}, Ph.D. Thesis, Harvard University, 1999.

\bibitem{Morschetal06}{\sc O. Morsch and M. Oberthaler}, {\em Dynamics of Bose-Einstein condensates in optical lattices},
Rev.\ Mod.\ Phys., 78 (2006), pp.\ 179--215.

\bibitem{oliva}{\sc J. Oliva}, {\em Density profile of the weakly interacting Bose gas confined in a potential well: Nonzero temperature},
Phys.\ Rev.\ B, 39 (1989), pp.\ 4197--4203.

\bibitem{stuart-book}{\sc G.~A. Pavliotis and A.~M. Stuart},
{\em Multiscale Methods: Averaging and Homogenization}, Springer, New York, NY, 2010.

\bibitem{Pitaevskii61}
{\sc L.~P. Pitaevskii}, {\em Vortex lines in an imperfect Bose gas},
Soviet Phys.\ JETP, 13 (1961), pp.\ 451--454.

\bibitem{Pitaevskii03}{\sc L. Pitaevskii and S. Stringari}, {\em Bose-Einstein Condensation}, Oxford Press, Oxford, UK, 2003.

\bibitem{popovfadeev}{\sc V.~N. Popov and L.~D. Faddeev}, {\em An approach to the theory of the low-temperature Bose gas}, Sov.\ Phys.\ JETP, 20 (1965), pp.\ 890--893.

\bibitem{Proukakis08}{\sc N.~P. Proukakis and B. Jackson},
{\em Finite-temperature models of Bose–Einstein condensation},
J.\ Phys.\ B: Mol.\ Opt.\ Phys., 41 (2008), 203002.

\bibitem{Solovej-summer}{\sc J.~P. Solovej}, {\em Many Body Quantum Mechanics}, Lecture Notes (2007); available at 

\quad http://www.mathematik.uni-muenchen.de/$\sim$sorensen/Lehre/SoSe2013/MQM2/skript.pdf \looseness=-1

\bibitem{spohn80}{\sc H. Spohn}, {\em Kinetic equations from Hamiltonian dynamics: Markovian limits},
Rev.\ Mod.\ Phys., 52 (1980), pp.\ 569--615.

\bibitem{Stenger99}{\sc J.~Stenger, S. Inouye, M.~R. Andrews, H.-J. Miesner, D.~M.
Stamper-Kurn, and W. Ketterle}, {\em Strongly enhanced inelastic collisions in a Bose-Einstein condensate near Feshbach resonances}, Phys.\ Rev.\ Lett., 82 (1999), pp.\ 2422--2425.

\bibitem{Wu98}{\sc T.~T. Wu}, {\em Bose-Einstein condensation in an external potential at zero temperature: General theory}, Phys.\ Rev.\ A, 58 (1998), pp.\ 1465--1474.

\bibitem{Wu61}
{\sc T.~T. Wu}, {\em Some nonequilibrium properties of a Bose system of hard
spheres at extremely low temperatures}, J.\ Math.\ Phys., 2 (1961), pp.\ 105--123.



\bibitem{Zarembaetal99}{\sc E. Zaremba, T. Nikuni, and A. Griffin}, {\em Dynamics of trapped Bose gases at finite temperatures},
J.\ Low Temp.\ Phys.,  116 (1999), pp.\ 277--345.



\end{thebibliography}
\end{document}